\documentclass{aa}

\usepackage{amsmath, amssymb, graphicx, txfonts}
\newcommand{\asr}{Adv. Space Res.}
\newcommand{\ag}{Ann. Geophys.}

\begin{document}
\title{Heavy Coronal Ions in the Heliosphere: I. Global Distribution of Charge-states of C, N, O, Mg, Si and S}
\author{S. Grzedzielski \inst{1} \and M.E. Wachowicz \inst{1} \and M. Bzowski \inst{1} \and V. Izmodenov \inst{2}}
\institute{Space Research Centre, Polish Academy of Sciences, Bartycka 18A, 00-716, Warsaw Poland \\
\and Lomonosov Moscow State University, Department of Mechanics and Mathematics \& Space Research Institute (IKI) Russian Academy of Sciences, Moscow, Russia}

\date{}

\abstract{}
{Investigate/Study de-charging of solar wind C, N, O, Mg, Si and S ions and assess fluxes of resulting ENA in the heliosphere.}
{ The model treats the heavy ions as test particles convected by (and in a particular case also diffusing through) a hydrodynamically calculated background plasma flow, from 1~AU to the termination shock (TS), to heliosheath (HS) and finally to heliospheric tail (HT). The ions undergo radiative and dielectronic recombinations, charge exchanges, photo- and electron impact ionizations with plasma particles, interstellar neutral atoms (calculated on a Monte-Carlo model) and solar photons.}
{Highly-charged heavy coronal ions flowing with the solar wind undergo successive de-ionizations, mainly in the heliosheath, leading to charge-states much lower than in the supersonic solar wind. If Coulomb scattering is the main ion energy loss mechanism, the end product of these deionizations are fluxes of ENA of $\sim 1$~keV/nucleon originating in the upwind heliosheath that for C, Mg, Si and S may constitute sources of pickup ions (PUI) significantly exceeding the interstellar supply.}
{ Discussed processes result in (i) distinct difference of ion charge q in the supersonic solar wind (approximately $q \geq +Z/2$, $Z$ = atomic number) compared to that in the HS (approximately $0 \leq  q \leq +Z/2)$), (ii) probable concentration of singly ionized atoms ($q = +1$) in the heliosheath towards the heliopause (HP) and in the HT, (iii) possible significant production of ENA in the HS offering natural explanation for production of PUI, and -- after acceleration at the TS -- anomalous cosmic rays (ACR) of species (like C, Mg, Si, S) unable to enter the heliospheric cavity from outside because of their total ionization in the local interstellar medium.}

\keywords{heavy ions -- ENA -- heliosphere -- PUI -- ACR}
\maketitle

\section{Introduction}
    The ions of C, N, O, Mg, Si and S leave the Sun as multiply ionized; typically more than 99\% of  them have charge-states $q > +4$ (von Steiger et al. 2000; Gloeckler et al.1998), resulting from very high ionization rates in the corona. These q-values are usually taken as  ``frozen'' over the solar wind ride through the inner heliosphere. However, freezing must evidently fail in the case of long residence times, i.e. for ions in the heliosheath (HS) and heliospheric tail (HT), the regions that constitute the main reservoir of heavy ions in the heliosphere. 

In the present paper (paper I) we examine the situation in detail by developing a global  model of the time evolution of charge-states of C, N, O, Mg, Si and S ions as the solar plasma flows from the corona to the termination shock (TS), to HS, and finally to HT. We show that when successive solar wind plasma elements fill out the heliosheath, the ions undergo -- mostly by electron capture from neutral interstellar H and He atoms -- a significant reshuffling of charge-states, while possibly retaining their initial energies of $\sim 1$~keV/n. This leads to a number of as yet unexplored consequences, like distinct differences in prevailing ionic charge-states between the supersonic solar wind and the heliosheath (Sec.~3.1, Table \ref{q}), dependence of spatial distribution of charge-states on the rate of thermalization of heavy ions in heliosheath plasmas (Sec.3.2), concentration of singly ionized atoms towards the heliopause (Sec.3.1, Fig.\ref{mapyiz}), and probable production in this layer of significant fluxes of ENA of $\sim 1$~keV/nucleon (Sec.~5, Fig.\ref{log}).

An interesting paper addressing related issues was recently published by Koutroumpa et al. (2007). In this paper expected charge-exchange induced soft X-ray and EUV emissions of the solar wind heavy ions were examined, taking into account detailed space- and time-dependent variability. However, the authors discuss essentially emissions due to ions in charge-states as they emerge from the solar corona (``primary'' ions as they call them). While being justified in the supersonic solar wind, their approach does not describe the deep ``reworking'' of ion charge-states in the heliosheath, which is the central topic of our paper and which -- as we show -- bears both on heavy neutral atoms and PUI (pickup ion) populations (and plausibly X- and EUV- emissions).

In a follow-up paper (paper II) we discuss the consequences of our modeling for heliospheric physics, concerning the experimental/observational detection of the predicted effects in XUV as well as in the form of fluxes of ENA, the possibility of  diagnosing the overall structure of the heliosheath, and the question of supplying seed ions (PUI) for the ACR populations of species with low-FIP (First Ionization Potential).

\section{The physical model}
\subsection{Test particle description of heavy ions}
We treat heavy ions as test particles carried by the general flow of interplanetary plasma, that undergo (binary) interactions with solar wind electrons, protons, with solar ionizing photons and with neutral atoms coming from interstellar space. We take into account radiative and dielectronic recombinations, impact ionizations, photoionizations and charge exchanges. A single interaction is assumed to alter the ionic charge $q$ by $\pm 1$, depending on the process. The most important, especially for multiply charged ions, is electron capture from neutral interstellar hydrogen. The time evolution along the flow line of the number $N\left(Z,A\right)^{+q}$ of ions with $0\leq q \leq Z$ (originating from atoms $Z,A$) and contained within a unit solar wind mass, can be described by  equations of the type
\begin{eqnarray}
\label{row1}
\frac{\mathrm{d}N_{(Z,A)}^{+q}}{\mathrm{d}t} &= &\sum (recombinations)+ \sum
    (ionizations)+ \nonumber \\ 
    & + &\sum (charge\  exchanges)
\end{eqnarray}
where the summations are over rates per unit mass per second for all relevant processes. To calculate the RHS of Eqs.\ref{row1} one should know, besides the relevant cross sections, the background solar plasma and neutral interstellar gas flows, i.e. the solar wind electron ($n_{e}$), proton ($n_{p}$) densities, the density distribution of interstellar neutral H and He ($n_{H}, n_{He})$ as well as the corresponding bulk flow velocities of the solar wind $(v_{sw})$ and interstellar gas $(v_{H})$ and the effective relative velocities of particles at collisions ($v_{rel}$) resulting from (local) particle velocity distribution functions.

\subsection{Flow of background plasma and neutral atoms}

The background flow of solar  plasma and neutral hydrogen atoms in supersonic solar wind, inner heliosheath and distant heliospheric tail was calculated basing on the time-independent, single-fluid, non-magnetic, gas-dynamical model for heliospheric proton-electron plasma coupled by mass, momentum and energy exchange with neutral interstellar hydrogen atoms as developed by Izmodenov and Alexashov (2003). In this self-consistent treatment the neutral H distribution was calculated kinetically (Monte-Carlo approach). The Sun as source of solar wind and ionizing photons is assumed to be spherically symmetric, with the wind speed of 450 km/s, Mach number 10 and $n_{p}=7$~cm~$^{-3}$ at Earth orbit. At infinity, a uniform interstellar flow of 25~km/s, with neutral hydrogen density $n_{H,LISM} = 0.2$~cm$^{-3}$, proton density $n_{p,{\mathrm{LISM}}} = 0.07$~cm$^{-3}$(= electron density) and temperature 6000~kelvin was assumed. To account for the presence of helium we took a simple model of a uniform He I substratum with atoms density $n_{\mathrm{He I}} = 0.015$~at./cm$^{3}$ (Gloeckler et al., 2004) flowing with velocity of 26.4~km/s (Witte 2004). Therefore our model disregards small scale features like the He I cavity and helium cone which are anyway of little consequence for the situation in the heliosheath.

Because of axial symmetry, all variables depend on the radial distance $r$ from the Sun and angle $\theta$ from the apex direction (i.e., direction of inflow of the local interstellar gas in the heliocentric frame). The evolution of spatial density of all charge-states for a species of atomic number Z was calculated by numerical integration of a set of $Z+1$ coupled ordinary, linear differential equations of type (1), in which the dependence of coefficients on the spatial coordinate along the flowline was given by the solutions of the combined hydrodynamic + Monte Carlo model mentioned above. The integration was carried along 180 flow lines, corresponding to initial (at Earth orbit) values of the angle $\theta$ counted from the apex direction equal to 1,2...180\degr. Using $\left|v_{sw}\right| \mathrm{d}t = \mathrm{d}s$, the time integration  can be transformed into a space integration along the curvilinear coordinate $s$ running along each of the flow lines. Such a procedure was performed for each of species separetely. In this way a complete spatial distribution of all $N(Z,A)^{+i}$ for every considered species could be obtained.

As long as the solar wind parcel moves supersonically between the Sun and the termination shock, the heavy ions can be thought to cool adiabatically like the background plasma and therefore stay in approximate thermal equilibrium with local plasma environment. In this case $v_{rel}$ for all interactions with electrons (radiative and dielectronic recombination, ionizing impacts) and for charge exchange reactions with protons was calculated assuming particle velocity distributions to be maxwellians corresponding to local (single-fluid) temperature. For heavy ion - neutral atom interactions $v_{rel}$ = solar wind bulk speed = 500~km/s was taken, as a rough compromise between the slow ($\sim$ near equatorial) and fast ($\sim$ high latitudes) solar wind streams.

Upon crossing the termination shock the proton-electron plasma on a single-fluid model heats up to about $\sim$ $ 10^{6}$ kelvin. However, there is no reason to assume the same single-fluid temperature applies to heavy ions. Even for relatively light ions like protons about one-fifth of their total population at quasi-perpendicular shocks flows downstream in form of a ring distribution in velocity space as immediately formed in the shock ramp (M{\"o}bius et al., 2001), i.e. unaffected by particle-wave coupling. Such tendency is even more plausible for various heavy ions that have no particular reason to be in resonance with the wave-field excited by the majority proton population. Rather, the downstream heavy ions may on the average approximately retain their range of upstream kinetic energies (about 0.5--3~keV/n for velocities 300--750 km/s), while undergoing pitch-angle scattering/reflection on shock structures and -- for a small fraction -- acceleration to energies much higher than in the upstream (Kucharek et al., 2006; Louarn et al., 2003). For strong shocks such behavior should lead to often considered assumption of downstream effective temperature of the heavy ions being proportional to ion mass. This is suggested by direct experimental evidence (Berdichevsky et al., 1997) from the Ulysses data on interplanetary shock crossings, when  downstream of the shock regions, characterized by a quasisteady plasma flux, values of $T(^{4}$He$^{2+})/T($H$^{+})$ and $T($O$^{6+})/T($H$^{+})$ are observed in the range 4.6 to 10.8 and 19 to 48, respectively (i.e. heating is even more than mass proportional). A less clear cut conclusion was recently drawn from a study of interplanetary shocks driven by the Coronal Mass Ejections  (SWICS spectrometer on the ACE spacecraft), (Korreck 2005; Korreck et al., 2006). Though heating seems to depend on several parameters like magnetic field angle, Mach number, plasma- $\beta$ and ion mass-to-charge, it appears, however, to be more efficient for  strong perpendicular shocks (like the TS).

\subsection{Isotropization versus thermalization}

Basing on these arguments we explore in the present model two limiting cases of heavy ion  behavior in the heliosheath depending on the assumed efficiency of coupling to the background plasma:
        
(i)~Isotropization. By this we call the situation when the fastest coupling mechanism is pitch-angle scattering into a velocity shell distribution by low-frequency electromagnetic waves excited by the initial velocity ring distribution, combined with scattering on a ``soup'' of coherent structures resulting from the shock transition (Alexandrova et al. 2004, 2006). The heavy ions isotropize momenta while preserving energy. Then, on a Coulomb time scale of energy exchange with background heliosheath plasma (and aided by inelastic  collisions with neutral H), the heavy ions cool down to the level of background temperatures. However, the typical time scale for Coulomb cooling of a $\sim 1$~keV/n ion on protons is  $10^{11}$~s for O$^{+8}$, and longer for lesser charges and higher masses. It is therefore much longer than the upwind heliosheath flow times of $10^{8} \sim 10^{9}$~s. As a consequence, the heliosheath plasma flowing along the flow lines should carry in its mist a (minor) population of heavy ions endowed with energies of the order of $\sim 1$~keV/n. These particles undergo, as the most important process,  electron capture collisions with neutral H and He. On top of that the heavy ions undergo all other mentioned binary processes, with rates (for electronic processes) corresponding to local ``hot'' maxwellian velocity distributions as governed by the temperature of the hydrodynamic single-fluid post-shock solution. We consider isotropization to be the most probable case for, at least, the bulk of the upwind heliosheath. Most of the results presented hereafter pertain to this situation (cf. Sec. 3.1).

(ii)~Thermalization. By this we understand the other extreme when upon the TS crossing the heavy ions adjust very quickly (say, within time $\tau \leqslant 10^{6} - 10^{7}$~s) their temperature to the temperature of the background plasma. This requires a very high rate of energy exchange between the heavies and protons, possibly by heavy ion resonant wave proton interactions. As the energy density $W_{w}$ of  waves induced by the heavies cannot exceed (Winske \& Gary 1986) $\sim$ one-half of energy density of the heavies themselves ($< 0.0005$ of the post-shock background energy density $W_{b}$), the shortest time $\tau$ behind the TS would be of the order of (Gary 1991):
\begin{equation}
\label{tau}
\tau \approx ( dominant\ resonant\ wave\ frequency)^{-1} \times
\frac{W_{b}}{W_{w}} \sim 10^6 {\mathrm{s}}
\end{equation}
for the extreme case of a saturated cyclotron turbulence peaked exactly at the gyrofrequencies of the dominant (at TS) O$^{+6}$ and O$^{+5}$ ions. In this estimate a post-shock magnetic field of 0.1~nT was assumed (Burlaga et al., 2005). There are no indications that such a turbulence does indeed prevail behind the TS and even if it did, other heavy ions with different mass-to-charge ratios might miss the peak of gyrofrequencies. We thus consider the ``thermalization'' case as rather less likely. However, for comparison and to get a better feeling of the situation calculations of such cases were also performed (cf. Sec. 3.2). The truth, probably, lies somewhere between (i) and (ii), plausibly more close to (i). Such calculations are now in progress.

\subsection{Cross sections and rates for relevant binary processes}

To describe the rates of binary processes affecting charge-states of heavy ions we tried to use the most reliable data. In particular radiative recombination rates were taken from (Aldrovandi \& Pequignot 1973; Verner \& Ferland 1996; Zatsarinny \& Gorczyca 2001)  and the dielectronic recombination rates following (Mazzotta et al., 1998; Zatsarinny \& Gorczyca 2001; Gorczyca et al., 2003). Electron impact ionization rates were taken into account for all  charge-states of considered species from the AMDIS data base and the photoionization rates for neutral and singly ionized ions, basing on compilations corresponding to average Sun. A significant effort was made to collect adequate cross sections for heavy ion electron capture from neutral H and He (Stancil et al.,1998, 1999; Kingdon \& Ferland 1996; Lin et al., 2005; Wang et al., 2003). 

\subsection{Initial values for heavy ion charge-states}

Integrations of Eqs.\ref{row1} for all six species were performed for various initial values of $N(Z,A)^{+i}$  based on in situ measurements by instruments MTOF/Celias on SOHO, SWICS on Ulysses and SWIMS on ACE (Bochsler et al., 2000; von Steiger et al., 2000; Reames, 1998). Values either taken at or reduced to Earth orbit (assuming $1/r^{2}$ dependence on heliocentric distance) were used. The presented final results are based on sets of relative initial $N(Z,A)^{+i}$  values averaged over the solar cycle, resulting from the SWICS measurements (Table \ref{Tabelka1}). The abundance data below $10^{-3}$ (empty fields in Table \ref{Tabelka1}) were confronted with independent data gathered in the MTOF/Celias experiment. In effective calculations the initial ratios of total number density of ions of a given species to solar wind proton density were taken from Raines et al. (2005) and von Steiger et al., (2000; Plate 2). The numerical values are as follows: O/H ratio = $5 \cdot 10^{-4}$,  C/O  = 0.67, N/O = 0.077, Mg/O = 0.145, Si/O = 0.146, S/O = 0.05.

\begin{table}[h]
\caption{Initial (1 AU) relative abundances of charge-states of heavy  ions (normalized to 1 for each species), based on in situ data from the SWICS instrument on Ulysses}
\label{Tabelka1}
\begin{center}
\begin{tabular}{|l|c|c|c|c|c|c|c|c|c|}
\hline
& +4& +5 & +6& +7& +8& +9& +10& +11& +12
\\
\hline
C&  0.6& 0.3& 0.1& & & & & &
\\
\hline
N &  & 0.4& 0.4&0.2& & &&&\\
\hline
O&&& 0.67& 0.3& 0.03&&&&\\
\hline
Mg&&& 0.02& 0.13& 0.15& 0.33& 0.37&&\\
\hline
Si&&& & 0.06& 0.17& 0.43& 0.18& 0.1& 0.06\\
\hline
S&&&&& 0.21& 0.29& 0.36& 0.14&\\
\hline
\end{tabular}
\end{center}
\end{table}

The relative abundances corresponding to experimentally undetectable levels of particular charge-states (empty fields in Tab. \ref{Tabelka1}) were assumed to be either $10^{-7}$ or $10^{-3}$. It was verified that results for charge-states that become dominant beyond the TS do not depend on these undetected values. On the whole it was found that results in the distant solar wind and heliosheath are only weakly sensitive to changes in the initial $N(Z,A)^{+i}$ values. The most important physical factor in the whole process turns out to be consecutive deionizations of heavy ions due to electron capture from neutral atoms. 

\section{Distribution of heavy ion charge-states in the heliosphere}
\subsection{Case isotropization}

\begin{figure*}
\centering
\includegraphics[width=17cm]{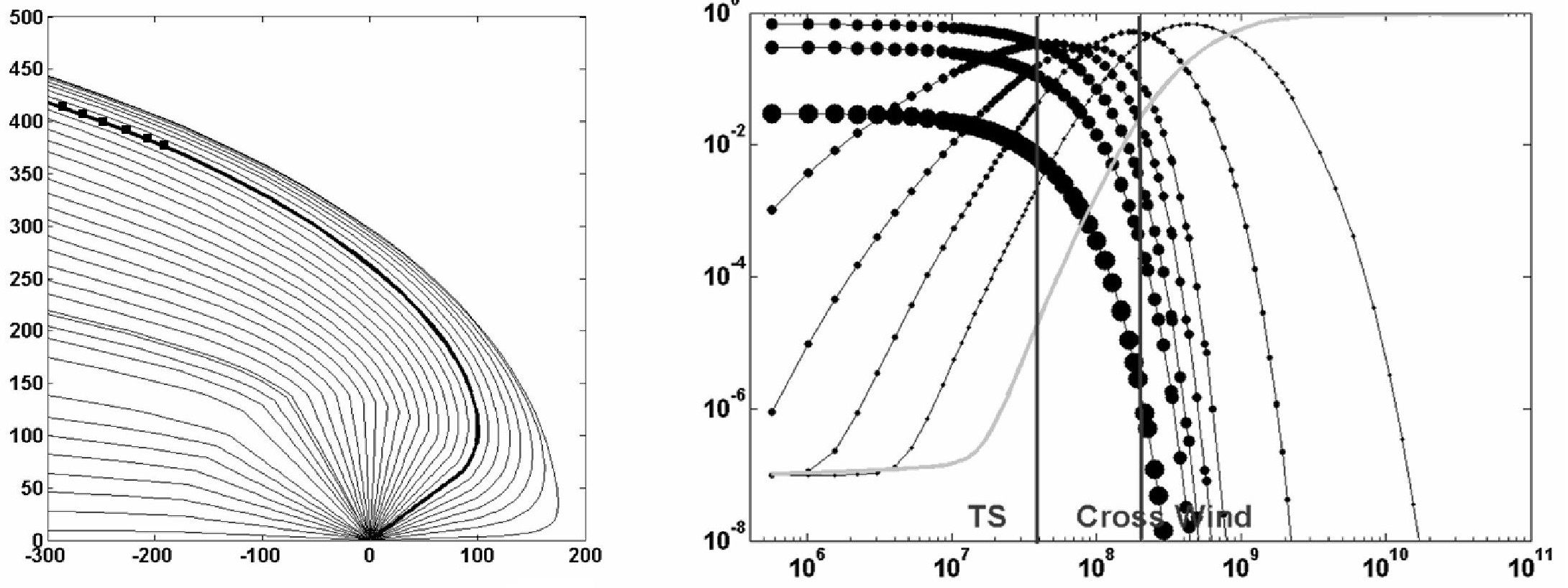}
\caption{The geometry of selected 30 (out of 180) flow lines in the  upwind and near tail heliosphere (left panel). As example (right panel) is shown the time evolution of relative abundances for all  charge-states of oxygen (i.e. $N_{(8,16)}^{+i}$ divided by the total number of O-ions per unit solarwind mass) along a flow line that starts $30\degr$ off apex direction. Decreasing size of dots corresponds to decreasing ion charge: largest dots -- O$^{+8}$, smallest dots -- O$^{+2}$, grey line -- O$^{+1}$. }
\label{Panelki}
\end{figure*}

The spatial distribution of all charge-states of  C , N,  O, Mg, Si, and S ions is constructed out of a grid of 180 solutions of Eq.\ref{row1} for each species, corresponding to individual flow lines starting at $\theta = 1,2...180\degr$ as described above. Each solution describes the time evolution (and, consequently, the spatial variability along the streamline) of the charge-states of considered elements. To describe the typical behavior  we show in Fig. \ref{Panelki} (left panel) the geometry of  selected 30 (out of 180) flow lines in the upwind and near tail heliosphere and, as example, (right panel) the time evolution of relative abundances for all charge-states of oxygen (i.e. $N_{(8,16)}^{+i}$ divided by the total number of O-ions per unit solar wind mass) along a flow line that starts 30\degr off apex direction. The case shown corresponds to isotropization. Vertical lines indicate times (in s) of termination shock and cross wind direction ($\theta = 90\degr$) passage. Decreasing size of black dots corresponds to decreasing ion charge (largest O$^{+8}$, smallest O$^{+2}$,grey line denotes O$^{+1}$). We call attention to the increasing importance  of low charge-states O-ions ($q = +1, +2, +3$) as plasma crosses the TS and approaches the cross wind (CW) direction. This is due to much longer plasma residence times in this region, about $\sim 10^{1}$ yrs as compared to $\sim 1$~year in the supersonic region. This acts in favor of electron capture processes, as reionization is very improbable for $q > +1$. About 70~yrs after leaving the Sun the bulk of oxygen ions is in form of O$^{+1}$. The section of the considered flow line for which O$^{+1}$ starts to dominate is indicated as black dots in the left panel. 

\begin{figure*}
\includegraphics[width=17cm]{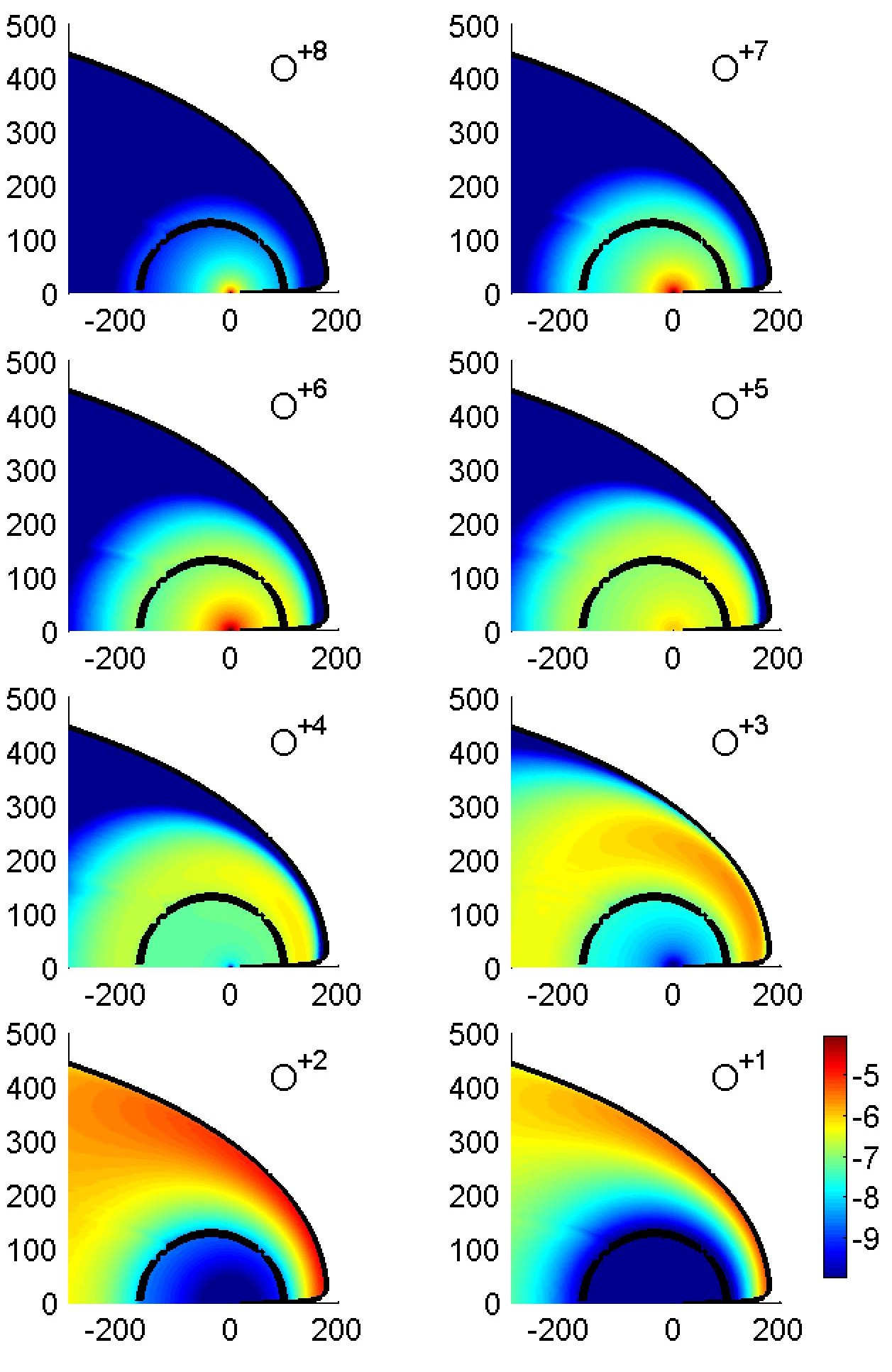}
\caption{ Heliospheric maps of density distributions (ions/cm$^{3})$ of oxygen ions in various ionization states under isotropization condition. The color coding in the right bottom corner corresponds to $\log[n_{\mathrm{ ion}} (\mathrm{in\, cm}^{-3})]$.}
\label{mapyiz}
\end{figure*}

One obtains qualitatively similar behavior for other flow lines starting into the upwind heliosphere. Note, however, the flow time scales from the TS to CW vary very significantly with $\theta$: from 70~years for $\theta = 3\degr$ to 1.6~years for $\theta = 80 \degr$. This means that the closer to apex direction starts a flow line, the sooner will it be dominated by the O$^{+1}$ ions. This tendency, combined with the topology of flow lines as shown in the left panel of Fig.\ref{Panelki}, means that the relative abundance of O$^{+1}$  will increase towards the heliopause all over the upwind heliosphere. The described behavior is illustrated in Fig.\ref{mapyiz}, which shows heliospheric maps of density distributions (ions/cm$^{3}$) of oxygen ions in various ionization states. Consecutive rows describe (left to right):  O$^{+8}$ -- O$^{+7}$, O$^{+6}$ -- O$^{+5}$, O$^{+4}$ -- O$^{+3}$, O$^{+2}$ -- O$^{+1}$ with common color coding for ion density (in cm$^{-3}$). Note a high density ridge appears for O$^{+3}$ beyond the termination shock. Finally, for O$^{+1}$ one obtains a strong density enhancement towards the heliopause.

Similar maps were obtained for all considered species, as shown in the Online Material. Two basic features are clearly prominent in all heavy ion distribution maps of the heliosphere:\\
(1) For all considered ions there is a definite difference in charge-states q in the supersonic solar wind as compared with the charge-states q in the heliosheath. For atoms of atomic number Z the divide lies around $q = +Z/2$. Typically, in the supersonic solar wind $q >\approx +Z/2$, while in the HS $q <\approx +Z/2$. Table \ref{q} shows the situation in detail.\\
(2)~Preferential concentration of singly-charged ions ($q=+1$) in certain regions of the heliosphere. Ions like C$^{+1}$, N$^{+1}$, O$^{+1}$, Si$^{+1}$ are most abundant on the upwind flanks of the heliosheath close to the HP, while Mg$^{+1}$ and S$^{+1}$ can be found on the distant flanks and in the heliotail.

\begin{table}[h]
\caption{Prevailing charge-states of heavy ions in the supersonic solar wind and heliosheath}
\label{q}
\begin{center}
\begin{tabular}{|l|l|l|}
\hline
Species& Supersonic solar wind (high q)&Heliosheath     (low q)\\
\hline
C&\scriptsize +4,+5,+6&\scriptsize +1,+2,+3,\\
\hline
N&\scriptsize+5,+6,+7&\scriptsize +1,+2,+3,+4,\\
\hline
O&\scriptsize+5,+6,+7,+8&\scriptsize +1,+2,+3,+4\\
\hline
Mg&\scriptsize+9,+10,+11,+12&\scriptsize +1,+2,+3,+4,+5,+6,+7,+8\\
\hline
Si&\scriptsize+10,+11,+12,+13,+14&\scriptsize +1,+2,+3,+4,+5,+6,+7,+8,+9\\
\hline
S&\scriptsize+8,+9,+10,+11,+12,+13,+14,+15,+16&\scriptsize+1,+2,+3,+4,+5,+6,+7\\
\hline
\end{tabular}
\end{center}
\end{table}

The above results indicate the upwind heliosheath should be far from homogeneous. On top of a (positive quasi-radial) gradient of ion density  (n-gradient$ > 0$), a (negative quasi-radial) gradient of ion charge-state (q-gradient $< 0$) is visible for most of the species. Note that while the n-gradient is a direct consequence of the background plasma distribution as calculated on hydrodynamic model, the q-gradient results from the interplay of flow line geometry and efficiency of individual binary processes altering the charge-states of ions. 

The typical density contrast between the maximum in $q=+1$ layer (which in most cases is lining up the heliopause) and the region adjacent to the termination shock is of the order of $\sim$ $10^{3}$, $\sim 5 \cdot 10^{4}$, $\sim 10^{5}$, $\sim 10^{5}$, $\sim 10^{6}$, for C$^{+1}$, N$^{+1}$, O$^{+1}$, Mg$^{+1}$, Si$^{+1}$. For O$^{+1}$ and Mg$^{+1}$ the $q=+1$ layer virtually lines up the heliopause; however, for C$^{+1}$, N$^{+1}$ and Si$^{+1}$ the maximum density in that layer is attained at distances, correspondingly, $\sim 2, 7$ and 3~AU from the heliopause (at $\theta = 30\degr$ off the apex direction). Existence of this relatively high density layer for singly ionized species is a direct consequence of the fact that heliosheath flow time scales are longest for the flow lines closest to the heliopause: this provides the ions with more chance for electron capture from neutrals. Differences between the positions of relative values of maxima for different species reflect particularities of individual cross sections. For instance, as can be seen from the maps, the regions of high density of Si$^{+1}$ and S$^{+1}$ are shifted towards the HT compared to the O$^{+1}$ density distribution. An overview  of  the heliospheric density distribution in the isotropization case for all charge-states of all considered species can be seen in the Online Material (Figs \ref{mapyCi}, \ref{mapyNi}, \ref{mapyMgi}, \ref{mapySii}, \ref{mapySi}). The format of these figures follows the format of Fig.\ref{mapyiz}.  

\subsection{Case thermalization}

The spatial distribution of all charge-states of C, N, O, Mg, Si, and S ions is obtained in a similar way as for the case of ``isotropization'' (cf. Sec. 3.1.). The main difference consists in different effective values for the relevant reaction rates, i.e. products of collision speed $v_{coll}$ times corresponding collision cross section $\sigma(v_{coll})$ for heavy ion interactions with neutral atoms and other plasma constituents. $v_{coll}$ is now determined mainly by the local single-fluid temperature as given by the hydrodynamic solution. As a consequence, for ion-neutral collisions instead of $v_{coll}\sim 500$~km/s, as in the case of isotropization, we have now values of tens of km/s only. For instance, for C-ions the typical  $v_{coll}$ values in the relatively dense, $\sim 50$ AU wide heliosheath layer adjacent to the heliopause amount now to 30-50 km/s in the heliosheath nose region $\theta = 0\degr$ and to $\sim 20-30$~km/s in the CW direction ($\theta = 90\degr$). As a result, the evolution of species by binary interactions is much slowed down compared to isotropization, while the hydrodynamic flow time scale remains unchanged. Because of this, the evolution of heavy ion charge-states along each of the flow lines is  now significantly retarded, i.e. successive de-ionizations of heavy ions occur much farther down the streamline. This translates into very different spatial distribution of particular charge-states when compared with isotropization.

\begin{figure*}
\centering
\includegraphics[width=17cm]{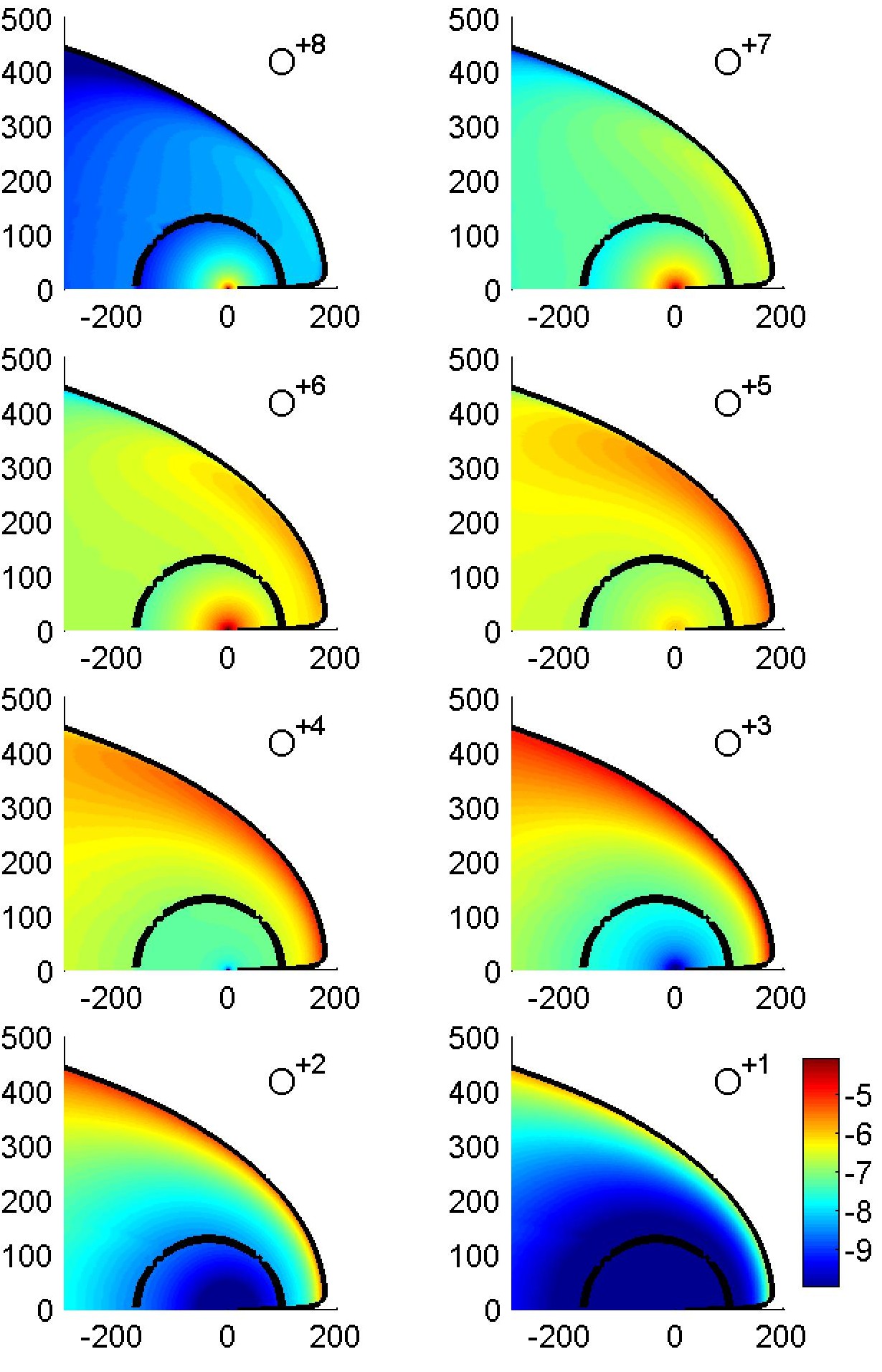}
\caption{ Heliospheric maps of density distributions (ions/cm$^{3}$) of oxygen ions in various ionization states under thermalization  condition (density coding as in Fig. \ref{mapyiz}).}
\label{mapyth}
\end{figure*}

To illustrate this effect we show  in Fig. \ref{mapyth} the density maps for O-ions in the case of thermalization in exactly the same format as for the case of isotropization in Fig. \ref{mapyiz}. The maps show the density distributions (ions/cm$^{3})$ of oxygen ions in various ionization states. Consecutive rows describe (left to right): O$^{+8}$ -- O$^{+7}$, O$^{+6}$ -- O$^{+5}$, O$^{+4}$ -- O$^{+3}$, O$^{+2}$ -- O$^{+1}$ with common color coding for ion density (in cm$^{-3}$).

One immediately notes important differences between the present ``thermalization'' case and the ``isotropization'' case shown in Fig. \ref{mapyiz}. For instance, such a high-charge state like O$^{+6}$ is now still very much present over the upwind heliosheath, and O$^{+5}$,O$^{+4}$ extend even well into the heliosheath tail area, while under isotropization these charge states were there virtually absent. A striking difference is also visible in the distributions of O$^{+3}$ ions. While under ``isotropization'' the density of these ions decreased sharply towards the heliopause, in the case of ``thermalization'' the reverse is true: the density attains maximum at the heliopause. Finally, concerning O$^{+1}$, one immediately recognizes that the amount of oxygen that was able to reach  this charge-state in the upwind heliosheath under thermalization is a tiny fraction of the corresponding amount converted to O$^{+1}$ under isotropization.

One obtains qualitatively similar differences between cases of isotropization and thermalization for all other considered ions. Maps of distribution for most of  charge-states for all considered ions can be found in  Online Material (Figs \ref{mapyCt}, \ref{mapyNt}, \ref{mapyMgt}, \ref{mapySit}, \ref{mapySt}). The format of these figures follows the format of Fig.\ref{mapyiz}. On the whole, it can be stated that in the case when isotropization holds the upwind heliosheath will be predominantly populated by the low charge-states while, when fast thermalization of heavy ions prevails, the ions will be in the high charge-states. This resembles the divide present in the isotropization case between the heavy ions in the supersonic solar wind and heliosheath (cf.Table~\ref{q}, Sec. 3.1). The precise meaning of low and high charge-states depends  in this context on species in question,  in function of the interplay between various reaction rates. It goes without saying that no higher charge-states  should appear in the heliosheath than those that are present in the solar corona. This comes out because of  vanishingly low probability of ionizing an already highly charged heavy ion.

The important conclusion one can infer from comparison of the``isotropization'' and ``thermalization'' cases is that observational determination of the prevailing charge-states in the upwind heliosheath should be indicative of the relative importance of plasma ``collective'' (i.e. waves, turbulence, etc.) thermalization processes in the heliosheath versus cooling by Coulomb scattering on background plasma and binary collisions with neutral atoms. Should low charge-states dominate, as we in fact expect, then collective effects would be of little importance and the heavy ions should stay hot well into the heliospheric tail. 

\section{Spatial diffusion}

A physical process, not mentioned hitherto, that may in principle affect the distribution of heavy ions in the heliosphere is spatial diffusion. Its effect over solar wind fast ride to the TS is obviously small, because of short time scale and heavy ion low energy as seen in the co-moving plasma frame. However, diffusion may carry ions away from the parent parcel of solar wind as the plasma moves relatively slowly along hydrodynamic flow lines in the heliosheath and, in addition, ions are hotter after the TS crossing. The distance, in coordinates co-moving with the fluid, covered in Brownian motion is of the order of $\mathrm{d} = \sqrt{\kappa_{ion}\,t_{hydr}}$, where $\kappa_{ion}$ is the diffusion coefficient for the heavy ions and $t_{hydr}$ is the hydrodynamic flow  time scale, counted from the TS. Obviously, diffusion is more important for faster thermal motion, i.e. it may be primarily of importance for the case of ``isotropization''. Unfortunately there are no direct data on diffusion in the heliosheath of ions of tens of keV energy. The Bohm diffusion coefficient for a 1~keV/n O$^{+5}$ ion in a $B = 0.1$~nT heliosheath magnetic field is $2 \cdot 10^{20}$~cm$^{2}$~s$^{-1}$. The Bohm diffusion is often considered to be a generous value for diffusion perpendicular to the magnetic field. However, one cannot exclude a priori even higher values, suggested by cosmic ray studies. To illustrate the possible consequences we provide here estimates for a very fast diffusion calculated using a diffusion coefficient taken as extrapolation to very low energies of the formula derived from global heliospheric distribution and solar cycle modulation of $\sim 0.1$-several GeV cosmic ray ions (Le Roux et al., 1996): 
\begin{equation}
\label{k}
\kappa_{\parallel} = \kappa_0 \frac{V}{c} F(P) {B_{e}}/{B}
\end{equation}

In this formula $\kappa_{\parallel}$ is the diffusion coefficient in the direction parallel to local the magnetic field $B$, $B_e$ is the field at 1~AU, $\kappa_{0}$ =$ 3.75 \cdot 10^{22}$~cm$^{2}$~s$^{-1}$, $v/c$ is the ion speed divided by the speed of light, $P$ is rigidity in GV, $F(P) = 0.4$ for the considered energy range. We assume that effective $\kappa_{ion}= 1/3 \cdot \kappa_{\parallel}$, and $B = 0.3$~nT in the sub-heliopause heliosheath plasma. As representative streamlines we take the lines starting at Earth with the off-apex angle $\theta = 30\degr$ and $10\degr$. The $t_{hydr}$ values corresponding to the flow from TS to the cross wind direction (CW, $\theta =90\degr$) are $7.5 \cdot 10^{8}$~s and $1.5 \cdot 10^{9}$~s. 

The resulting heavy ion diffusion off the hydrodynamic flow line is estimated below for O$^{+2}$ both for ``isotropization'' and ``thermalization''. In the latter case, for $T \sim 4 \cdot 10^{5}$~kelvin in the sub-heliopause plasma, $d \sim 3-5$~AU and is therefore rather small compared to heliosheath spatial scales. This suggests that were ``thermalization'' the proper description of heavy ions' thermodynamic state, a purely hydrodynamic model, as described in Sec.~3.2 would suffice. 

However, in the case of ``isotropization'' ion energies in plasma frame are $\sim 1$~keV/n and diffusion may no longer be negligible, especially at low $\theta$. For the mentioned streamlines the diffusive displacement over the heliosheath flow time scale amounts to $\sim 15-25$~AU. This means that details of solutions presented for ``isotropization'' (Sec.~3.1) in the upwind heliosphere may in reality become smeared out. To qualitatively assess the magnitude of possible effects we develop in the present section a simplified, time-independent, spherically symmetric, convective-diffusive description of the heavy ion flow. Results obtained with this approach suggest that even diffusion as fast as the one extrapolated from cosmic ray studies will not invalidate the main results obtained under the axisymmetric hydrodynamical model used as basis in the present paper.
 
The heavy ions are again treated as test particles carried by background plasma and interacting with a stream of neutral atoms entering the heliosphere cavity through the heliopause. As before, the ions undergo all binary processes as described by Eqs (\ref{row1}). This time, however, they are also allowed to diffuse through plasma with a diffusion coefficient $\kappa$ as given by Eq.(\ref{k}). The situation is commonly described by a cosmic-ray-type transport equation (Jokipii 1987). In our case the cosmic ray particles are replaced by the heavy ions and we consider only the total pressure $p_{hi}(r)$ of the heavy ion gas without attempting to describe possible evolution of the momentum distribution function (Drury \&  Voelk 1981). In spherical symmetry the transport equation for $p_{hi}(r)$ takes then the form (a separate equation for each ion): 
\begin{equation}
\label{transport}
\frac{1}{r^2}  \frac{\mathrm{d}}{\mathrm{d}r} {\left[\frac{r^2}{\gamma - 1}  \left(\gamma  p_{hi}  v - \kappa  \frac{\mathrm{d}p_{hi}}{\mathrm{d}r}\right)\right]} - v  \frac{\mathrm{d}p_{hi}}{\mathrm{d}r} = Q,
\end{equation}
where $\gamma$ is the adiabatic exponent for the heavy ion gas (assumed = 5/3) and $Q$ describes the loss process due to neutralization of singly charged ions by electron capture from neutral interstellar atoms. 

The background plasma is supposed to enter the HS through a spherical TS placed at heliocentric distance $r_{TS} = 106.9$~AU with a  purely radial hydrodynamic speed $v_{TS} = 115.121$~km/s on the downstream side. From the TS the plasma flows radially outwards with velocity $v(r)$ given by: 
\begin{equation}
\label{r}
v(r) = 115.121\cdot\left(0.6374- 4.7973 \cdot 10^{15}/r + 8.6518 \cdot 10^{30}/r^2\right),
\end{equation}
where $v$ is in km/s and $r$ in cm. The values of $r_{TS}, v_{TS}$, and the formula (\ref{r}) for $v(r)$, were chosen as weighted averages (over a $90\degr$ wide sector of the upwind heliosphere centered at the apex direction) of the corresponding TS distances and radial velocity components taken from the hydrodynamic axisymmetric model of Izmodenov \& Alexashov (2003). The heliocentric distance of the heliopause $(r_{HP})$ was determined by the requirement that the average (hydrodynamic) residence time, in the considered sector, of the background plasma in the heliosheath under the spherically symmetric diffusive model be the same as the corresponding average residence time in the axisymmetric hydrodynamic non-diffusive model, which is equal to $5 \cdot 10^{8}$~s. In this way the diffusive model allocates for all relevant physical processes the same (on the average) time scale to operate as in the case of hydrodynamic model. Under this condition the heliopause distance in the diffusive model is placed at $r_{HP} = 178.9$~AU and the final radial velocity is quite small, $v(r_{HP}) = 5.87$~km/s. 

Solutions of Eqs (\ref{row1}) and (\ref{transport}) with approximations (\ref{k}) and (\ref{r}) yield the distribution of heavy ion population under assumed radial convection-diffusion. Appropriate equations for each species were effectively solved with the following boundary conditions:\\
(1)The flux of heavy ions introduced into the (considered sector of) heliosheath at $r_{TS}$ corresponds to the flux carried (on the average) by the hydrodynamic flow,\\
(2) The heavy ion pressure at the HP vanishes ($p_{hi} = 0$), because of ion free escape into the external medium due to the expected high value of $\kappa$ in the local interstellar gas as compared with the heliosphere values (increase by 2 or 3 orders of magnitude). The high external values of $\kappa$ are suggested by appropriate formulae for $\kappa$  as function of particle rigidity in the interstellar medium (Axford 1981; Moskalenko et al., 2001) when extrapolated to the very low energy domain considered in the present context.

\begin{figure}
\centering
\includegraphics[width=8cm]{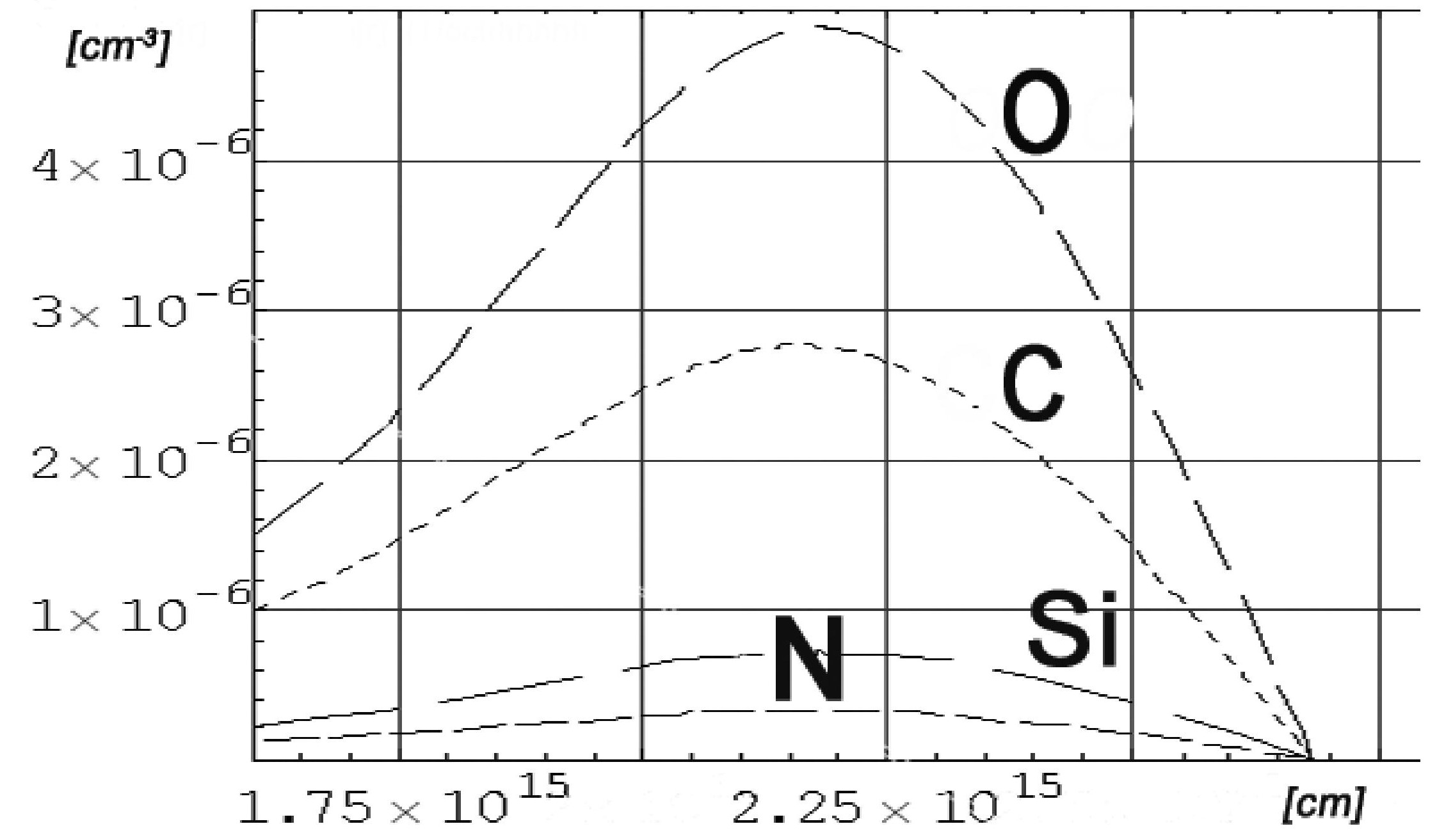}
\caption{Heavy ion densities for diffusive, spherically symmetric solution. The left border of the diagram corresponds to the termination shock position $r_{\mathrm{TS}} = 106.9$~AU. Heliopause position $r_{\mathrm{HP}} = 178.9$~AU corresponds to vanishing density.}
\label{dyfuzja}
\end{figure}
 
Results of this diffusive model indicate that indeed diffusion may alter the density contrast compared to purely hydrodynamic axisymmetric isotropization solution described in Sec. 3.1. The situation is summarized in Fig.\ref{dyfuzja} which shows the radial distribution of the total ion pressure ($p_{hi}$) for each species. $p_{hi}(r)$ first increases due to slowdown of the convective transport as implied by Eq. (\ref{r}), then attains a maximum somewhere midway between the TS and HP, and finally decreases towards the HP. This decrease is due both to increased electron capture neutralization of $q = +1$ ions, resulting from the slowness of convection in that region, and to the diffusive escape into the interstellar medium. It was checked that the obtained density distribution is insensitive to the details of the transition between the low $\kappa$-values inside the heliopause to the high $\kappa$-values outside, provided the increase is by more than a factor of 30, which is satisfied for the Axford and Moskalenko et al. formulae mentioned above. 

It is evident from Fig. \ref{dyfuzja} the heavy ion density gradient (n-gradient) under the diffusive isotropization  model is much less than for the hydrodynamical model. The density contrast between the midway maximum and the post shock values ranges from 2.7 for C to 3.3 for Si. However, the total ENA production by neutralization of ions is much the same as in the purely hydrodynamic case (cf. Sec. 3.1). 

\section{Production of energetic neutral atoms (ENA) in the heliosheath}
\subsection{Expected fluxes of ENA at 1~AU}

\begin{figure}
\centering
\includegraphics[width=8cm]{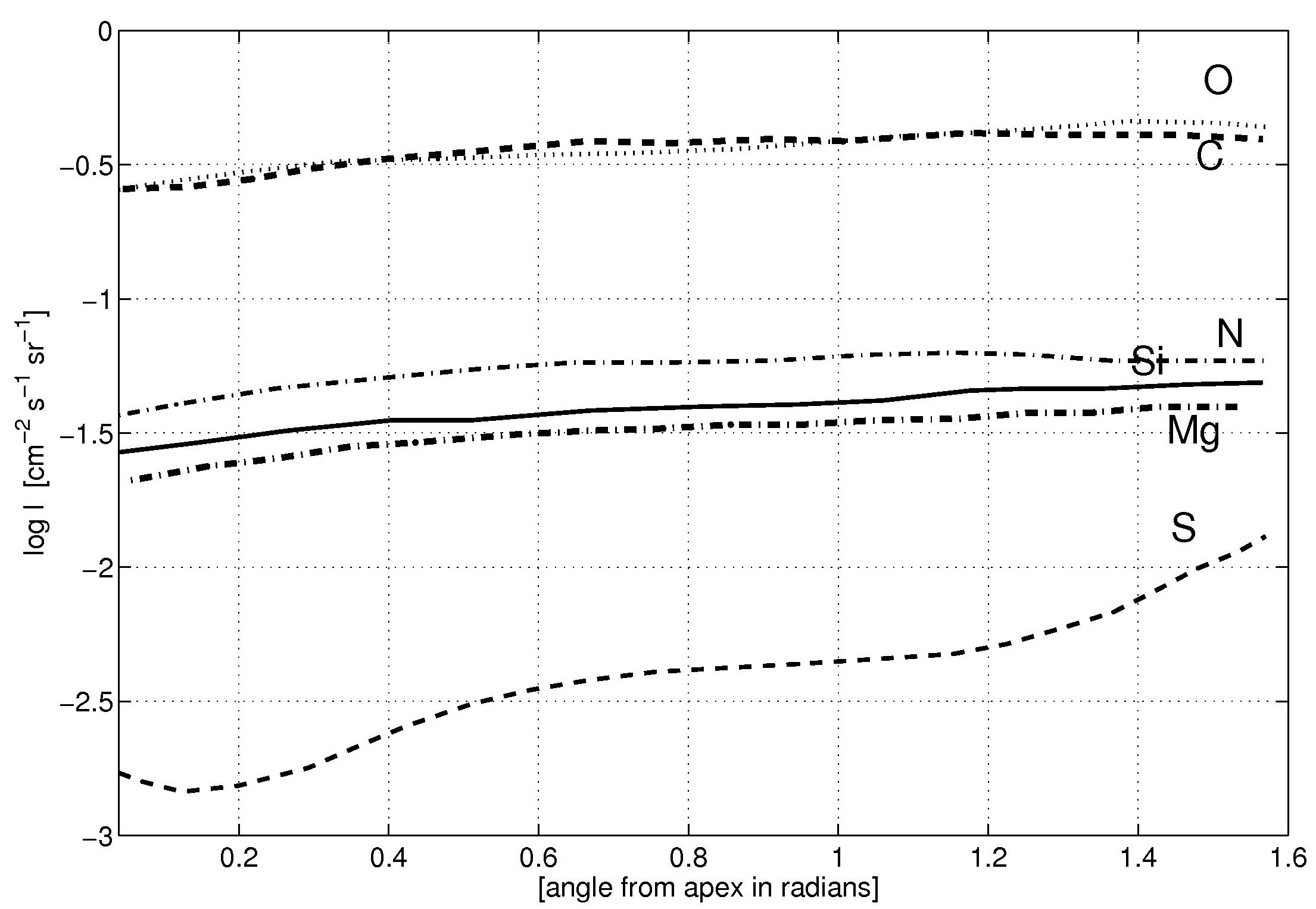}
\caption{Logarithm of intensities of ENA (cm$^{-2}$~s$^{-1}$~sr$^{-1}$) emitted by heliosheath plasma, as function of angular distance of the LOS from the apex direction (radians), seen from the Sun if no losses intervened. Curves from top to bottom (right side of diagram) correspond to: O, C, N, Si, Mg, S. }
\label{log}
\end{figure}

Singly ionized ions produce neutral atoms by charge exchange with interstellar H (and He). In the case of ``isotropization'' the resulting neutral atoms will inherit the $\sim 1$~keV/n energies (ENAs). The intensity $I$ (in atoms~cm$^{-2}$~s$^{-1}$~sr$^{-1}$) of ENA fluxes for a particular species is given by an integral over the LOS (line-of-sight) of the source function:
\begin{equation}
\label{natezenie}
I(\theta) = \frac{1}{4 \pi} \int n_{+1}(n_{H}\sigma_{H} + n_{He}\sigma_{He})\, v_{coll}\, \mathrm{d}r,
\end{equation}
where $n_{+1} $ denotes the density of singly charged ions, $n_{H}, n_{He}$ are the densities of neutral hydrogen and helium atoms, and $\sigma _{H}$,$ \sigma_{He}$ the corresponding cross sections for electron capture (a common $v_{coll}$ suffices in view of small speeds of H and He atoms).  Intensities $I(\theta)$ resulting from integration of expression (\ref{natezenie}) for our model values over various LOS emanating from the Sun are given for all considered species as function of angle $\theta$ from the apex direction in Fig. \ref{log}. 

\begin{table}
\caption{Survival probabilities for 1~keV/nucleon ENA flight from the heliopause to 1~AU} 
\label{ena} 
\centering
\begin{tabular}{c c c c c c c} 
\hline
species  & C & N & O & Mg & Si & S \\
survival probability($\%$)& 58 & 75& 73& 69& 34& 64 \\
\hline 
\end{tabular}
\end{table}

These intensities are not corrected for the losses that ENA will undergo during their flight from the heliosheath. We calculated the losses for a more realistic situation, when the observer is displaced from the Sun by 1~AU along the LOS. The calculations include photoionization losses assumed to vary $\sim 1/r^{2}$, charge exchange losses with the supersonic solar wind, varying also $ \sim 1/r^{2}$, losses induced by electron impact ionization in the supersonic solar wind for electron temperature $T_{e}$ varying as in Marsch et al. (1989), charge exchange losses in the heliosheath for plasma density simulated by a two-step function (half of the path with $n_{p} = n_{e} = 0.005$~cm$^{-3}$, the other half with $0.04$~cm$^{-3}$), and electron impact ionization losses in the heliosheath assuming $T_{e} = 10^{6}$ kelvin. The resulting correction factors (ENA survival probability over flight to 1~AU) to be applied to intensities $I(\theta)$ shown in Fig.\ref{log} are given in the Table~\ref{ena}.
\begin{table}[h]
\caption{Comparison of heliosheath supply of PUI with interstellar supply of PUI (only the upwind section of the heliosheath is assumed to produce ENA).}
\label{duza}
\begin{center}
\begin{tabular}{|c|c|c|c|}
\hline
 Species &  Heliosheath     &  I/S neutrals & Heliosheath/interstellar    \\
 ENA PUIs&  ENA PUIs (g/s)  & PUIs (g/s)    &  source ratio\\
\hline
C       &                  $4.5 \cdot 10^6$             & $1.8 \cdot 10^5$ & 25 \\
\hline
N                 &        $4 \cdot 10^5 $    &  $6.3 \cdot 10^7$&$6 \cdot 10^{-3}$ \\
\hline
O           &              $3.7 \cdot 10^6 $        &  $6.7 \cdot 10^8$& $5 \cdot 10^{-3}$\\
\hline
Mg          &           $ 1.8 \cdot 10^5$&        $5.1 \cdot 10^3$&35\\
\hline
 Si          &             $1.5 \cdot 10^6$& $7.4 \cdot 10^1$&$2\cdot 10^4$\\
\hline
 S             &            $6.5 \cdot 10^4$&  $5.2 \cdot 10^2$&125\\
\hline
\end{tabular}
\end{center}
\end{table}

The ENA intensities from Fig.\ref{log} corrected for survival probabilities as in Tab.\ref{ena} suggest that in the case of ``isotropization'' the expected fluxes may attain 0.1~at./(cm$^{2}$~s~sr) for C and O and about $\sim 10^{-2}$~at./(cm$^{2}$~s~sr) for  N, Mg and Si. One may hope that fluxes of $\sim 0.1$~at./(cm$^{2}$~s~ sr) could be within the reach of a dedicated experiment such as the forthcoming NASA SMEX mission IBEX. If successful, such measurements would provide a method to directly diagnose the velocity distribution function of heavy ions in the heliosheath. Investigation of variation of ENA intensities over the sky could inform about possible asymmetries in the shape of the heliopause, be it due to external magnetic field or to non-uniform distribution of the surrounding interstellar plasma. We stress, however, that such interesting possibilities are very much contingent on the fulfillment of conditions for ``isotropization''.

It is also important to note that were diffusion significant for isotropization (cf. Sec.4) the overall intensity of ENA fluxes would not change much, though details of angular dependence as shown in Fig. \ref{log} might look different. In the case of ``thermalization'', presence of measurable fluxes of neutral atoms from the heliosheath seems improbable on three accounts: very low particle energies, high losses over flight to 1~AU, displacement of sources from the upwind heliosheath to (more distant) tail regions. A more detailed discussion of the opportunities to use possible detection of ENA fluxes for the diagnostics of the strucuture of heliosphere will be presented in Paper II. 

\subsection{Pickup ions from ENA as seed particles for anomalous cosmic rays}

ENAs entering the supersonic solar wind region between the Sun and TS constitute an additional, compared to neutral interstellar atoms, source of pickup ions (PUI). The importance of this source relative to interstellar supply can be assessed by comparing the fraction of the total flux of ENA  crossing the TS from downstream that become ionized in the supersonic solar wind, with the corresponding ionized fraction of the total (parallel) flux of interstellar neutral atoms. Such a comparison was made separately for each species, and we took into account the same loss processes as discussed in Sec. 5.1. Concerning the geometry,  we assumed for simplicity a spherical configuration with the supersonic solar wind constituting for the neutral atoms a circular target with radius equal to the TS radius (=~106.9~AU) and all ENA sources, assumed equidistant from the Sun, contained within a narrow emitting layer of radius 178.9~AU lining up from inside the heliopause. Actually, we took only sources contained in the upwind half of the heliosheath, as our modeling did not consistently include the tail section of the heliosphere. We recall that such a restriction of ENA sources to a narrow layer is typical in the ``isotropization'' case (cf. Sec. 3.1). Numerical values for both radii were taken identical as in Sec. 4. Note on the other hand, should diffusion be important (Sec. 4.), the sources of ENA would be distributed all over the heliosheath and the contribution of ENA to production of PUI would increase due to larger geometrical factor of the solar wind target.

The flux of interstellar atoms impinging on the heliosphere was estimated basing on results by Slavin \& Frisch (2007) (their model 26). In their modeling they undertook a detailed analysis of ionization conditions and related abundances of various species in the interstellar gas at solar location, taking into account both local in situ data on fluxes of interstellar atoms, column densities of gas towards nearby stars and data on ionizing radiations from stars and the Local Bubble. In particular Slavin and Frisch analyzed issues related to plausible significant ionization gradient in the LIC, and reinterpreted in this context the interstellar abundances. For instance, they infer local interstellar C seems to be overabundant compared to the solar standard and Mg and Si highly depleted by deposition onto interstellar grains. In our calculation we used, basing on the  model 26, the following values for the neutral fractions of atoms of the six considered species (Slavin \& Frisch, Table~6):  C -- $2.69 \cdot 10^{-4}$,  N -- $0.720$, O -- $0.814$, Mg -- $1.98 \cdot 10^{-3}$, Si -- $4.21 \cdot 10^{-5}$, S -- $6.47 \cdot 10^{-5}$. For the purpose of our model we used [O]/[H] = 331 ppm, hydrogen ionization degree 0.224 (both values also from their Table~6) and assumed interstellar neutral hydrogen density at solar location equal to 0.15~at.~cm$^{-3}$.

The results of our calculations on the relative importance of PUI created by the ENA fluxes expected according to present modeling as compared with the interstellar supply are shown in Tab.\ref{duza}. It is evident that for species like N and O that are thought to be largely neutral in the local interstellar gas the relative contribution to PUI production in supersonic solar wind by heliospheric ENA is insignificant. However, for the low-FIP species like C, Mg, Si, S, which should be virtually totally ionized in front of the heliosphere (Slavin \& Frisch, 2007), the heliosheath PUI supply resulting from our modeling can be orders of magnitude more significant than the interstellar. It is worth noting that our estimate of carbon PUI supply by deionization in the heliosheath exceeds the total carbon PUI supply from all other so-called ``inner'' and  ``outer'' sources, like outgassing of comets, grain sputtering, solar wind neutralization on grains, that are invoked (Schwadron et al., 2002)  to explain the PUI seed ions for the observed ACR carbon. As values in Table \ref{duza} indicate, the heliosheath ENA may constitute even more  attractive candidates for PUI in the case of other low-FIP species accelerated to ACR energies. It is therefore tempting to speculate that deionization of heavy solar ions in the heliosheath, combined with subsequent ENA drift into and ionization by the supersonic solar wind provides the necessary mechanism for production of seed particles for heliospheric ACR populations of most, if not all, low-FIP species present in ACR spectra. A detailed discussion of this question we present in Paper~II.

\section{Final remarks and conclusions}

The gist of the present paper lies in the observation that for the presently estimated densities of neutral interstellar atoms at heliosphere's peripheries, the time scale for complete deionization (by electron capture from the neutrals) of heavy solar ions convected with the solar wind may be comparable with the plasma flow time in the heliosheath ($\sim 10^{8} - 10^{9}$~s). The important proviso is that heavy ions lose their $\sim 1$~keV/nucleon energy slow enough to secure high collision rates. This last condition, in turn, is satisfied if the cooling of heavy ions is due primarily to Coulomb scattering (time scale of $\sim 10^{11} s$) on the relatively cold background (bulk) plasma, i.e. when no energy equilibration by collective plasma processes is operative in the post-termination shock solar wind (we call this case ``isotropization'' to stress that ions randomize their velocities while preserving energies, Sec.2.3). As we show in our modeling, the concurrence of the above conditions would result in definite predictions concerning the state of plasma populations in the outer heliosphere:\\
1.~The charge-states of heavy ions in the heliosheath should be much lower than in the supersonic solar wind, implying possible opportunities for detection by detailed analysis of soft X-ray and EUV emissions (Sec.3.1) (this method could also be of interest for study of astrospheres around nearby stars). This issue is discussed for some simple cases in Paper II.\\
2.~Neutralization of singly-charged heavy ions concentrated predominantly very close to the upwind flanks of the heliopause should give rise to fluxes of ENA, that could -- at least for carbon and oxygen -- be within reach of a dedicated instrument (Sec.3.1 and 5.1);\\
3.~ENA produced in the vicinity of heliopause will drift all over the heliosphere and upon (re)entering the supersonic solar wind and (re)ionization therein will provide sources of PUI, which for the considered low-FIP species (C, Mg, Si, S) exceed other possible sources of ACR seed populations (Sec. 5.2, Table \ref{duza}).\\

Predictions 2. and 3. are strongly contingent on the assumption of isotropization. Should these effects be not detected in the future, or be they marginal, would strongly suggest that equilibration of heavy ion energies with the background flow proceeds fast enough to imply efficient coupling of heavies to bulk plasma via wave excitation etc. In this way such negative outcome would also provide a way of probing the state of heliosheath plasmas. Finally, let us note that with no isotropization (i.e. under thermalization, Sec. 3.2) the effects mentioned under 1. above do not disappear altogether, but are present in a modified way: instead of a rather sharp charge-states difference between the supersonic solar wind, one obtains gradual spatial shifts in the heliosheath and near tail of  high density ``islands'' of particular charge-states. As already  mentioned before, a number of the mentioned issues will be addressed in a follow-up paper now in preparation (Paper~II).

\onlfig{6}{
\begin{figure*}
\centering
\includegraphics[width=17cm]{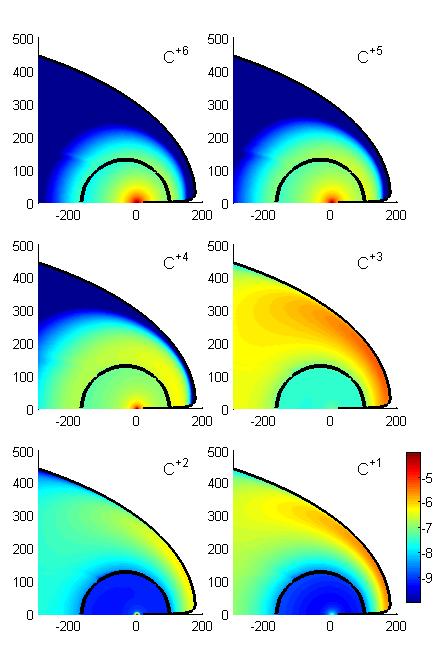}
\caption{ Heliospheric maps of density distributions (ions/cm$^{3}$) of carbon ions in various ionization states under isotropization condition (density coding as in Fig. \ref{mapyiz}).}
\label{mapyCi}
\end{figure*}
}

\onlfig{7}{
\begin{figure*}
\centering
\includegraphics[width=17cm]{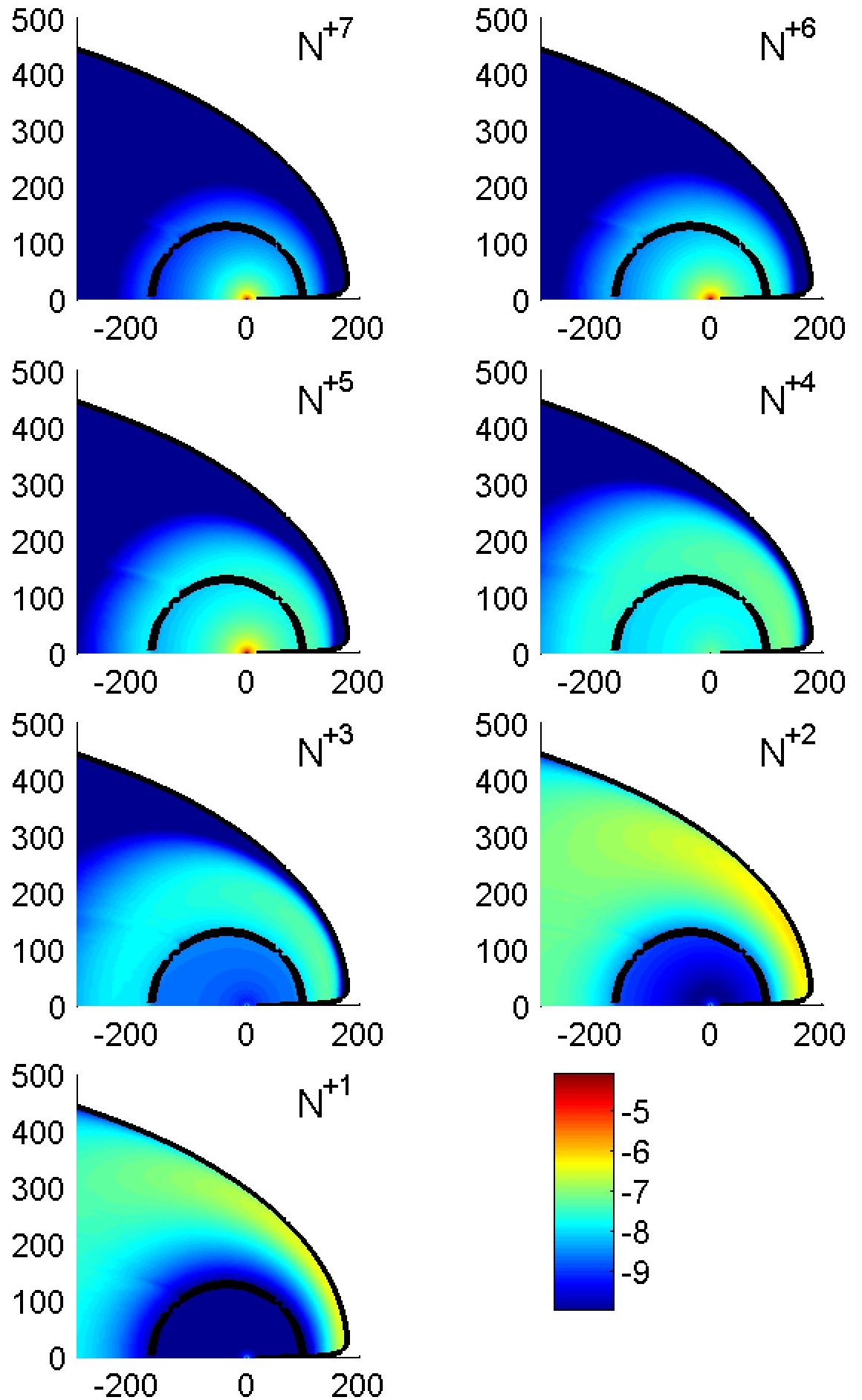}
\caption{ Heliospheric maps of density distributions (ions/cm$^{3}$) of nitrogen ions in various ionization states under isotropization condition (density coding as in Fig. \ref{mapyiz}).}
\label{mapyNi}
\end{figure*}
}
\onlfig{8}{
\begin{figure*}
\centering
\includegraphics[width=17cm]{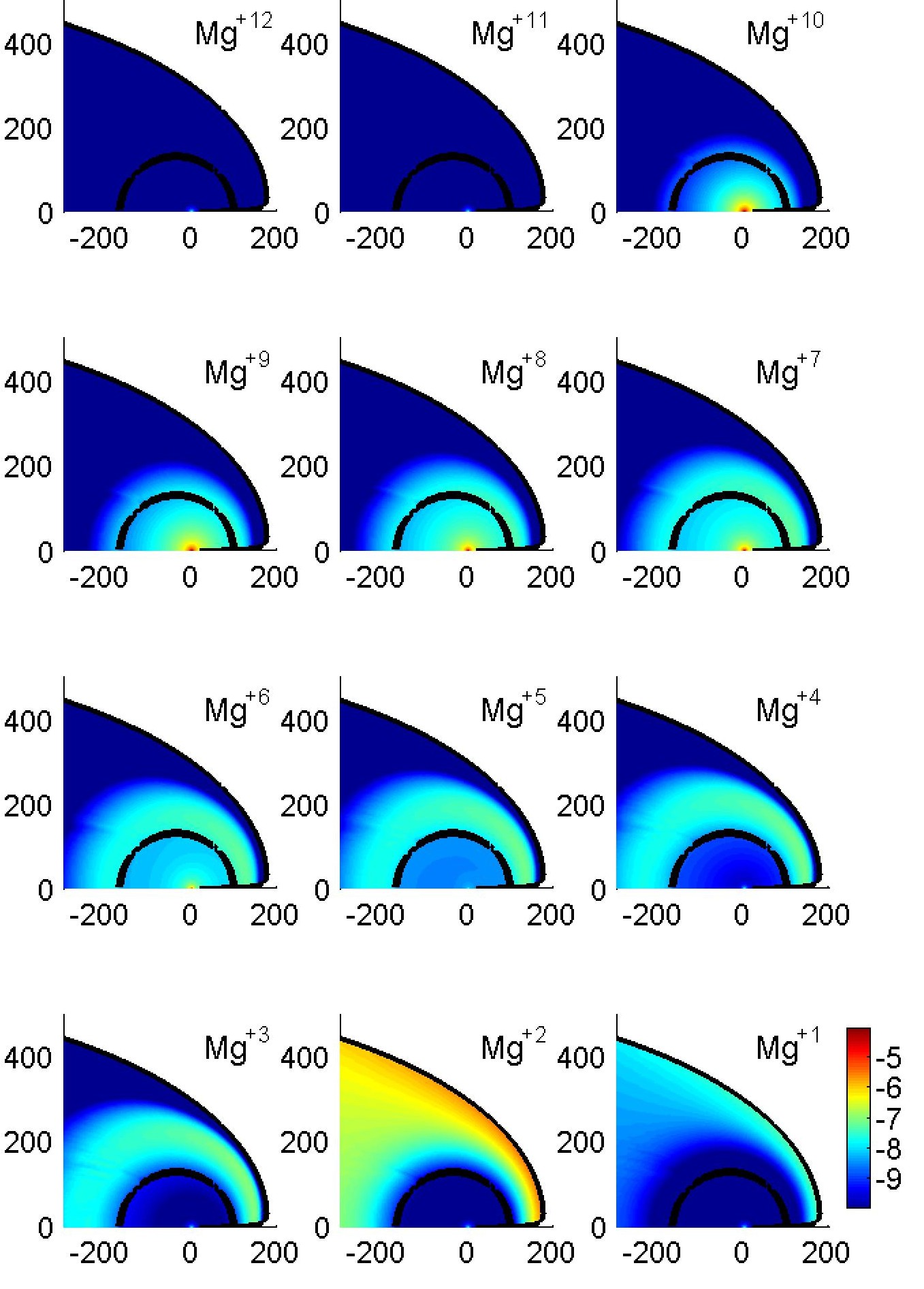}
\caption{ Heliospheric maps of density distributions (ions/cm$^{3}$) of magnesium ions in various ionization states under isotropization condition (density coding as in Fig. \ref{mapyiz}).}
\label{mapyMgi}
\end{figure*}
}

\onlfig{9}{
\begin{figure*}
\centering
\includegraphics[width=17cm]{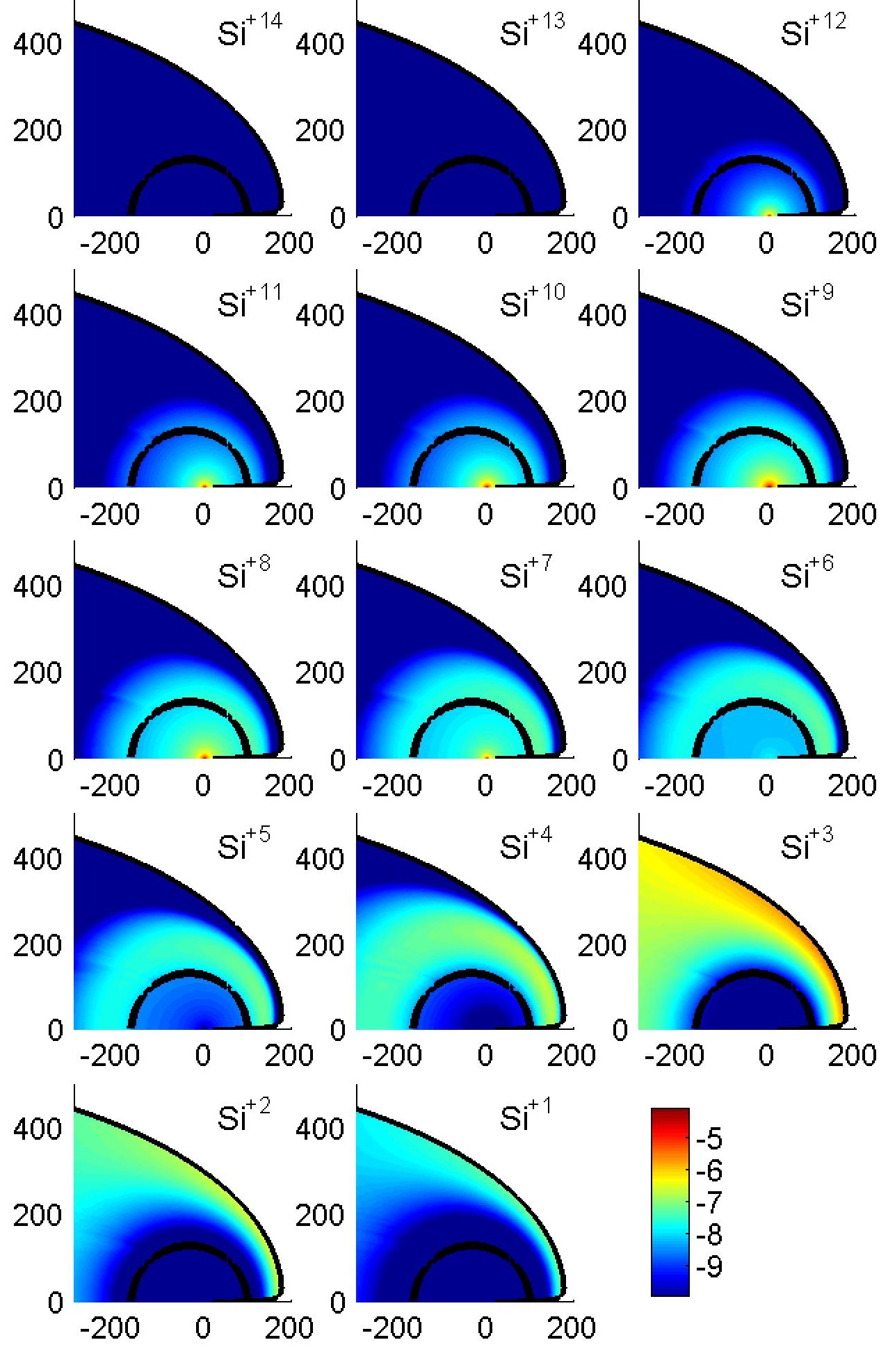}
\caption{ Heliospheric maps of density distributions (ions/cm$^{3}$) of silicon ions in various ionization states under isotropization condition (density coding as in Fig. \ref{mapyiz}).}
\label{mapySii}
\end{figure*}
}

\onlfig{10}{
\begin{figure*}
\centering
\includegraphics[width=17cm]{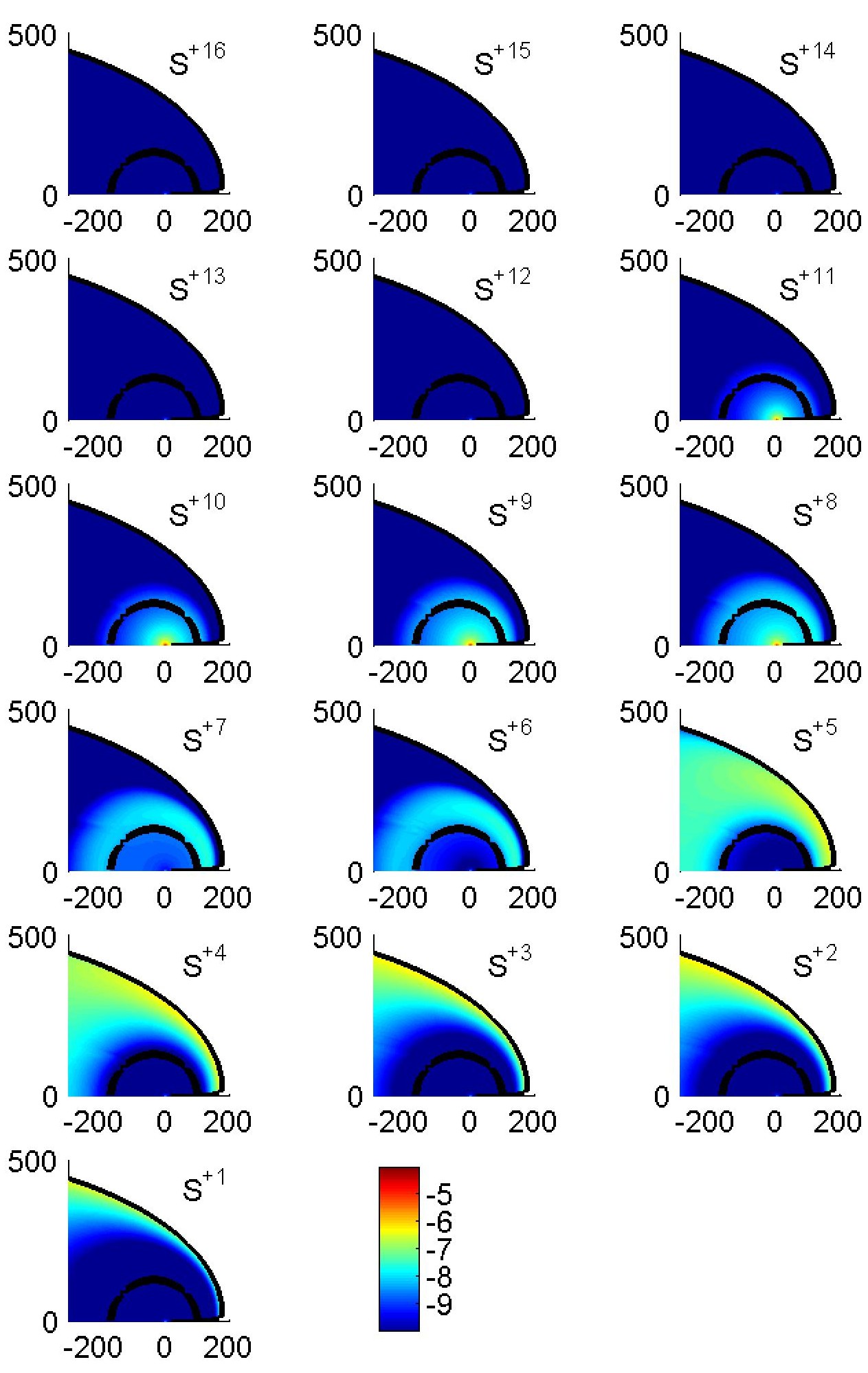}
\caption{ Heliospheric maps of density distributions (ions/cm$^{3}$) of sulfur ions in various ionization states under isotropization condition (density coding as in Fig. \ref{mapyiz}).}
\label{mapySi}
\end{figure*}
}
\onlfig{11}{
\begin{figure*}
\centering
\includegraphics[width=17cm]{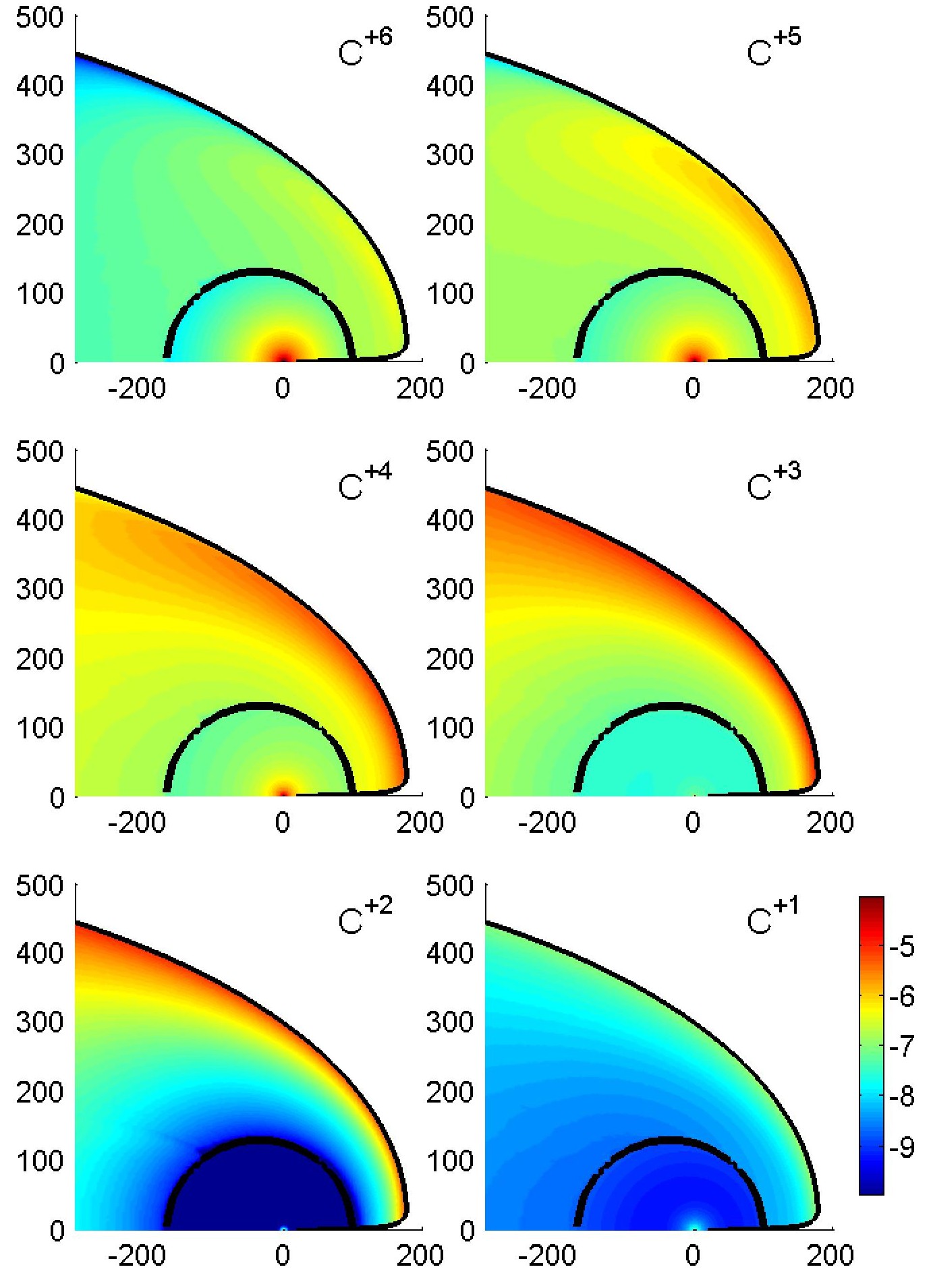}
\caption{ Heliospheric maps of density distributions (ions/cm$^{3}$) of carbon ions in various ionization states under thermalization  condition (density coding as in Fig. \ref{mapyiz}).}
\label{mapyCt}
\end{figure*}
}

\onlfig{12}{
\begin{figure*}
\centering
\includegraphics[width=17cm]{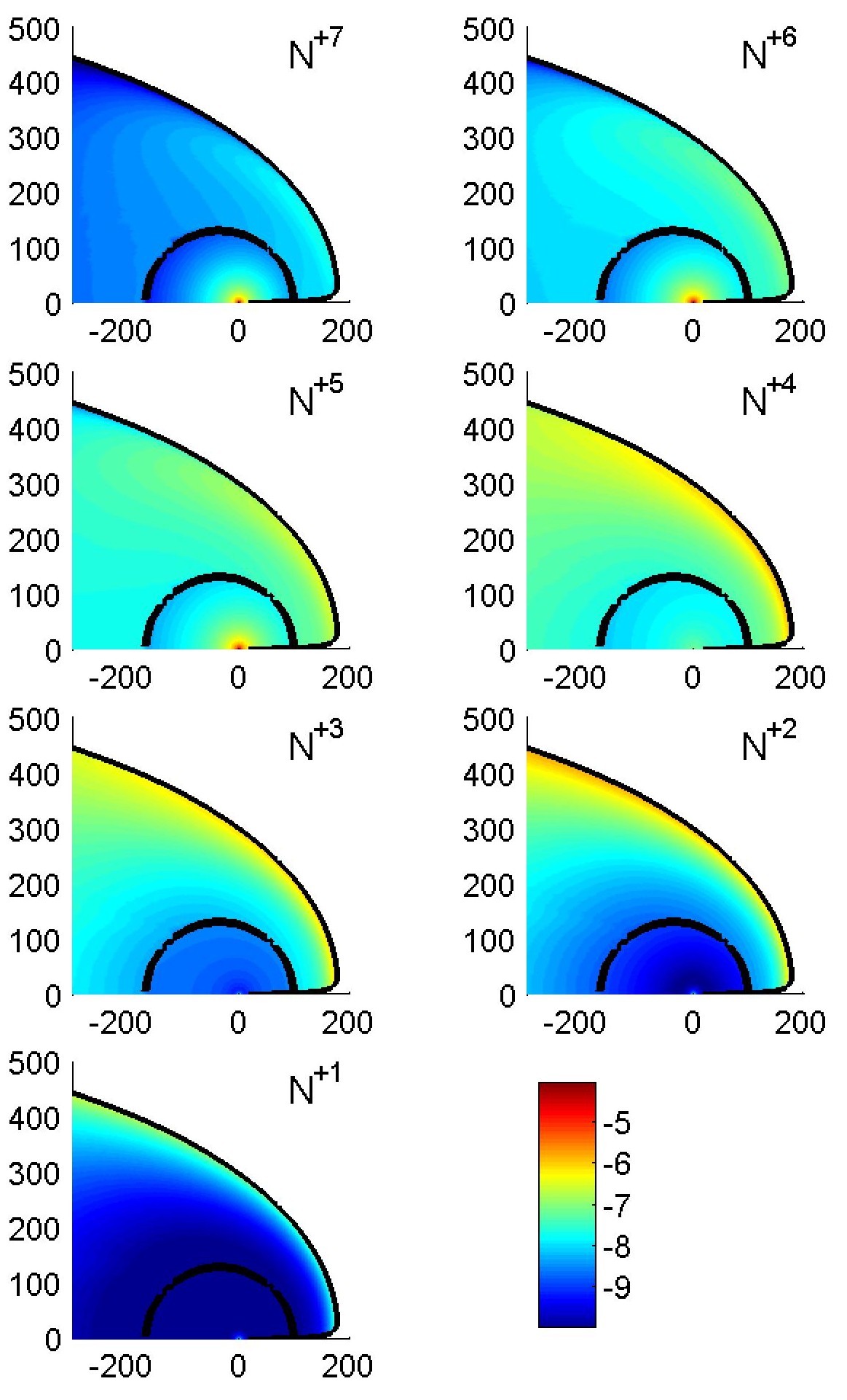}
\caption{ Heliospheric maps of density distributions (ions/cm$^{3}$) of nitrogen ions in various ionization states under thermalization  condition (density coding as in Fig. \ref{mapyiz}).}
\label{mapyNt}
\end{figure*}
}

\onlfig{13}{
\begin{figure*}
\centering
\includegraphics[width=17cm]{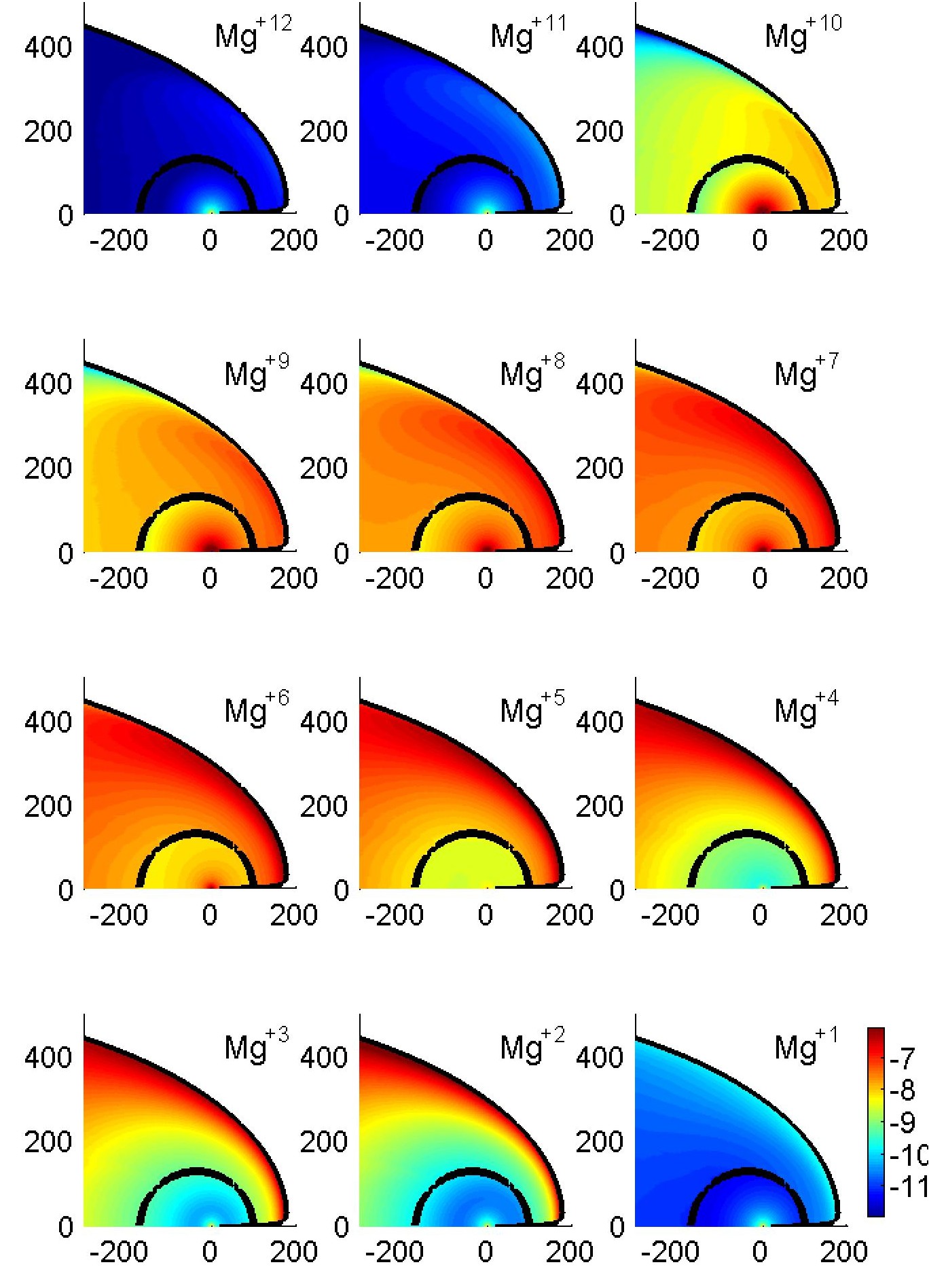}
\caption{ Heliospheric maps of density distributions (ions/cm$^{3}$) of magnesium ions in various ionization states under thermalization  condition (density coding as in Fig. \ref{mapyiz}).}
\label{mapyMgt}
\end{figure*}
}

\onlfig{14}{
\begin{figure*}
\centering
\includegraphics[width=17cm]{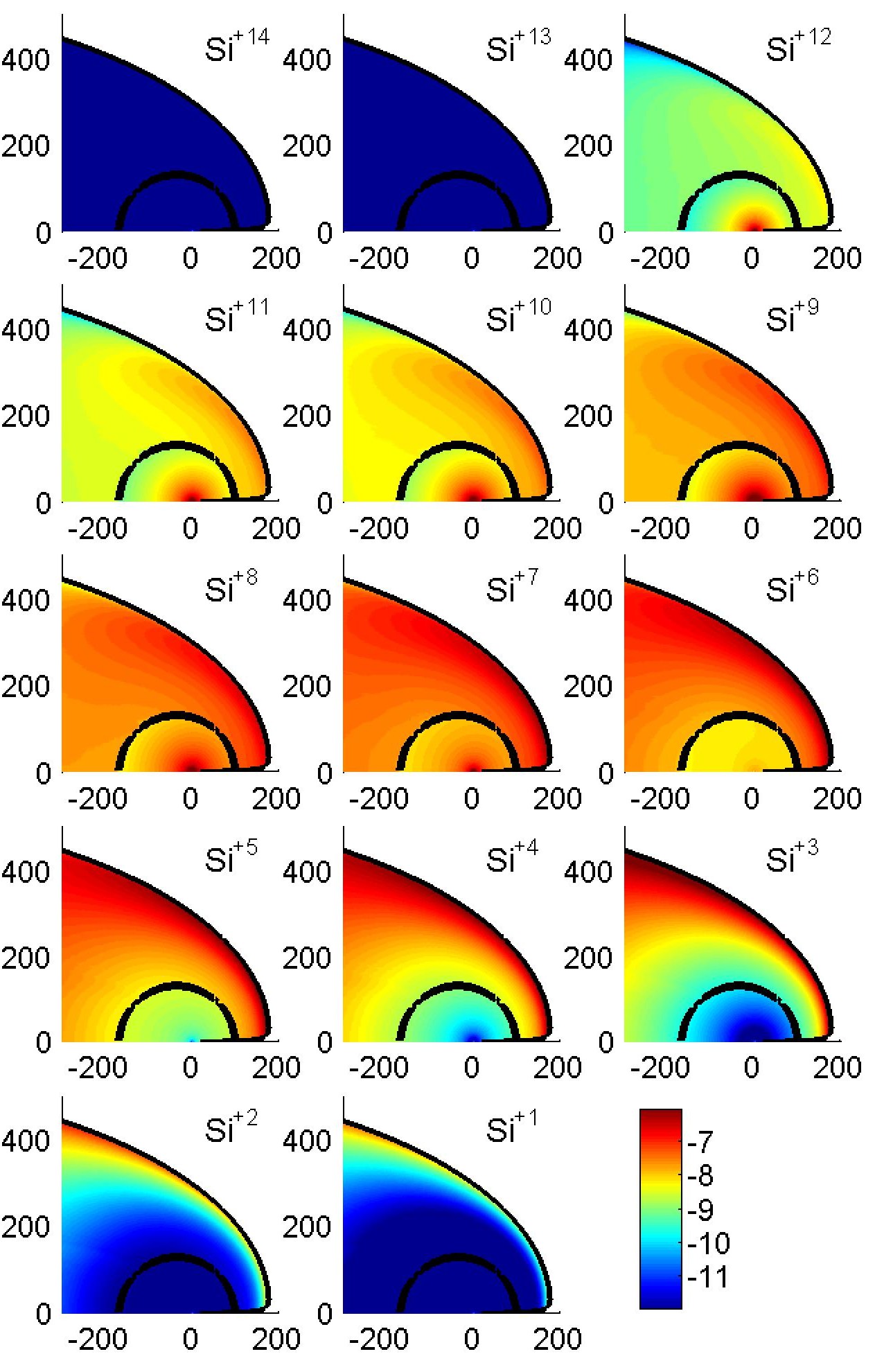}
\caption{ Heliospheric maps of density distributions (ions/cm$^{3}$) of silicon ions in various ionization states under thermalization  condition (density coding as in Fig. \ref{mapyiz}).}
\label{mapySit}
\end{figure*}
}

\onlfig{15}{
\begin{figure*}
\centering
\includegraphics[width=17cm]{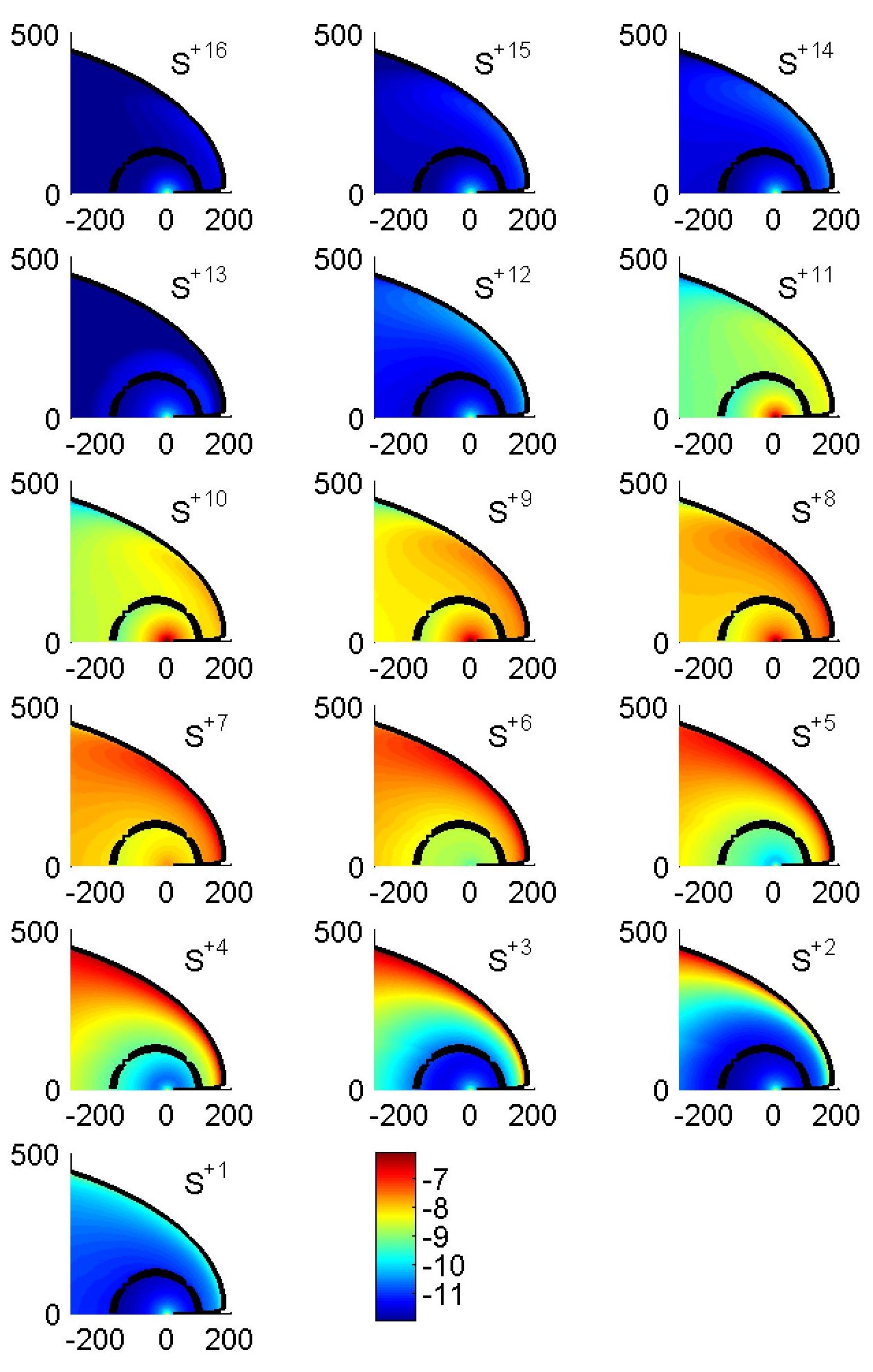}
\caption{ Heliospheric maps of density distributions (ions/cm$^{3}$) of sulfur ions in various ionization states under thermalization  condition (density coding as in Fig. \ref{mapyiz}).}
\label{mapySt}
\end{figure*}
}
\begin{acknowledgements}

That research has been supported by the Polish MSRiT grants 1P03D00927 and N522 002 31/0902.

\end{acknowledgements}

\end{document}